# An *in vivo* Comparison Study Between Strontium Nanoparticles and rhBMP2

Giulia Montagna[1,2†], Francesco Cristofaro[1†], Lorenzo Fassina[2], Giovanna Bruni[3], Lucia Cucca[4], Alejandro Kochen[5], Paola Divieti Pajevic[5], Beth Bragdon[6*], Livia Visai[1,7*] and Louis Gerstenfeld[6]

[1] Department of Molecular Medicine (DMM), Center for Health Technologies (CHT), UdR INSTM, University of Pavia, Pavia, Italy, [2] Department of Electrical, Computer and Biomedical Engineering, University of Pavia, Pavia, Italy, [3] C.S.G.I. Department of Chemistry, Physical-Chemistry Section, University of Pavia, Pavia, Italy, [4] Department of Chemistry, University of Pavia, Pavia, Italy, [5] Department of Translational Dental Medicine, Goldman School of Dental Medicine, Boston University, Boston, MA, United States, [6] Department of Orthopaedic Surgery, Boston University School of Medicine, Boston, MA, United States, [7] Department of Occupational Medicine, Toxicology and Environmental Risks, Istituti Clinici Scientifici Maugeri, IRCCS, Pavia, Italy

The osteoinductive property of strontium was repeatedly proven in the last decades. Compelling *in vitro* data demonstrated that strontium hydroxyapatite nanoparticles exert a dual action, by promoting osteoblasts-driven matrix secretion and inhibiting osteoclasts-driven matrix resorption. Recombinant human bone morphogenetic protein 2 (rhBMP2) is a powerful osteoinductive biologic, used for the treatment of vertebral fractures and critically-sized bone defects. Although effective, the use of rhBMP2 has limitations due its recombinant morphogen nature. In this study, we examined the comparison between two osteoinductive agents: rhBMP2 and the innovative strontium-substituted hydroxyapatite nanoparticles. To test their effectiveness, we independently loaded Gelfoam sponges with the two osteoinductive agents and used the sponges as agent-carriers. Gelfoam are FDA-approved biodegradable medical devices used as delivery system for musculoskeletal defects. Their porous structure and spongy morphology make them attractive in orthopedic field. The abiotic characterization of the loaded sponges, involving ion release pattern and structure investigation, was followed by *in vivo* implantation onto the periosteum of healthy mice and comparison of the effects induced by each implant was performed. Abiotic analysis demonstrated that strontium was continuously released from the sponges over 28 days with a pattern similar to rhBMP2. Histological observations and gene expression analysis showed stronger endochondral ossification elicited by strontium compared to rhBMP2. Osteoclast activity was more inhibited by strontium than by rhBMP2. These results demonstrated the use of sponges loaded with strontium nanoparticles as potential bone grafts might provide better outcomes for complex fractures. Strontium nanoparticles are a novel and effective non-biologic treatment for bone injuries and can be used as novel powerful therapeutics for bone regeneration.

# INTRODUCTION

Bone tissue constitutes the rigid scaffold that is the skeleton, which provides structural support for vertebrates and confers protection to the most delicate vital organs. The extraordinary regenerative capability of bone tissue to repair and heal without the formation of a fibrotic scar was recognized by Imhotep (2630–2611 BC) and Hippocrates (460–370 BC) (Lanza and Vegetti, 1974). However, in pathological conditions such as osteoporosis, sever bone trauma or critically-sized bone defects this process can be hindered, leading to delayed healing and/or non-union fractures (Holroyd et al., 2008; Sozen et al., 2017). These unsolicited outcomes constitute an often underestimated burden for our society, in term of quality of life, recovery time and costs for the healthcare system (Gentleman et al., 2010; Marcus, 2007).

Bone tissue engineering focuses on the creation of devices capable of providing physical support, activating bone forming cells (concepts, respectively, defined as osteoconductivity and osteoinductivity) with the aim of assisting and accelerating osteogenesis (Dennis et al., 2015; Kowalczewski and Saul, 2018). During the last two decades *in vitro* and *in vivo* studies were designed to identify novel scaffolding systems for a topical and controlled delivery of osteoinductive agents (Landi et al., 2007; Visai et al., 2017; Chandran et al., 2018). Many of them focused on the employment of recombinant human bone morphogenic protein 2 (rhBMP2) as a powerful osteoinductive agent (Wang et al., 1990; Buza and Einhorn, 2016). A great amount of effort has been placed in investigating rhBMP2 suitability as treatment for conditions such as spinal fractures and critically-sized bone defects (Noshi et al., 2001; Boix et al., 2005). Several studies highlighted the benefits of using this recombinant morphogen, but also its side effects (Wang et al., 1990; Noshi et al., 2001). Local side effects (such as insufficient or excessive bone formation, heterotopic bone formation, infections and inflammatory responses) as well as systemic ones (cancer in a relatively low percentage of cases) were identified in several independent studies (Boraiah et al., 2009; Latzman et al., 2010; Hoffmann et al., 2013; Woo, 2013; Poon et al., 2016). Compelling experimental results showed the efficacy of strontium cations, as osteoinductive and anti-osteoporotic agents (Rohnke et al., 2016; Carmo et al., 2018; Li et al., 2018). Strontium modulates bone remodeling by enhancing bone formation and suppressing bone resorption (Chattopadhyay et al., 2006; Takaoka et al., 2010; Yang et al., 2011; Saidak and Marie, 2012; Tian et al., 2014) although, to date, the molecular and cellular mechanisms of strontium activity remain partially elusive. Physical-chemical properties of strontium-substituted hydroxyapatite nanoparticles in combination with calcium hydroxyapatite nanoparticles and their *in vitro* osteoinductivity were assessed in our previous works (Frasnelli et al., 2016; Visai et al., 2017; Cristofaro et al., 2019).

In this work we utilized new nanotechnology to design a novel nanomaterial that is absent of recombinant morphogen yet still possess bone healing properties exerted by strontium to be used as an application for orthopedic surgery. This nanomaterial is a combination of strontium-substituted and calcium hydroxyapatite nanoparticles which was delivered *in vivo* by Gelfoam sponges, an FDA-approved collagen-based sponge commonly used as hemostatic application on bleeding surfaces (Pharmacia and Upjohn Company and Pfizer, 2017). The combination between the porous structure and molecular composition made it attractive in the orthopedic field and suitable either for morphogens or for ceramics delivery (Rohanizadeh et al., 2008; Giorgi et al., 2017).

The aim of this work is to compare the *in vivo* osteogenesis of strontium hydroxyapatite nanoparticles (SrHAn) to calcium hydroxyapatite nanoparticles and rhBMP2 (HAn-BMP2) using Gelfoam sponges as carriers with both treatments. An extensive characterization was performed including an abiotic study on the implant structure, as well as analysis of ions and rhBMP2 release patterns. The *in vivo* study was performed, on a healthy, bone damage-free mouse model to analyze the effectiveness of the described implants in targeting bone tissue. The sponges were implanted adjacent to the periosteum and gene expression of markers of osteocytes and OCs maintenance, MSCs recruitment, osteogenic and chondrogenic differentiation were evaluated, in order to analyze the responses at the periosteal bone. The experimental scheme is shown in **Figure 1**, with the comparison between SrHAn and HAn-BMP2. Sponges loaded with only calcium hydroxyapatite nanoparticles (HAn) were used as the negative control. In the study we included also the stability assessment of lyophilized samples, as this lyophilization could be suitable for the packaging and distribution of the implants.

# MATERIALS AND METHODS

## Standard Sponges Loading Protocol

Gelfoam sponges (Gelfoam; Pfizer®, NY, USA) were cut into 1 cm$^2$. The sponge absorbance capacity of 200 µL was defined as the volume of water completely absorbed by each sponge. The sponges were loaded with a HCl-acidified MilliQ water solution (pH 5) containing 30% (w/v) of calcium hydroxyapatite nanoparticles (HAn) or with a MilliQ water solution containing 27% (w/v) of HAn and 3% (w/v) of strontium hydroxyapatite nanoparticles (SrHAn). For abiotic characterization unloaded sponges were hydrated with acidified MilliQ and defined as the control (CTRL). For rhBMP2 loaded-sponges, 3 µg of human recombinant BMP2 (R&D Systems) were added to the HAn suspension before the sponge absorption phase (HAn-BMP2). Acidified MilliQ was used to increase nanoparticles solubility, buffer pH suspension (otherwise alkaline) and to prevent rhBMP2 precipitation (Friess et al., 1999; Luca et al., 2010). The CTRL, HAn, HAn-BMP2 and SrHAn samples were incubated at room temperature (RT) in a 24-well plate for 15 min after loading to allow the volume to be adsorbed. Following 30 min of incubation at 37°C, in a humidified and controlled atmosphere, with 5% CO$_2$, the wells were filled with 800 µL of acidified MilliQ.

---

**Abbreviations:** ECM, extracellular matrix; OBs, osteoblasts; OCs, osteoclasts; CaSR, calcium sensing receptor HAn, calcium hydroxyapatite nanoparticles; SrHAn, mixture of 90% calcium hydroxyapatite nanoparticles and 10% strontium hydroxyapatite nanoparticles.

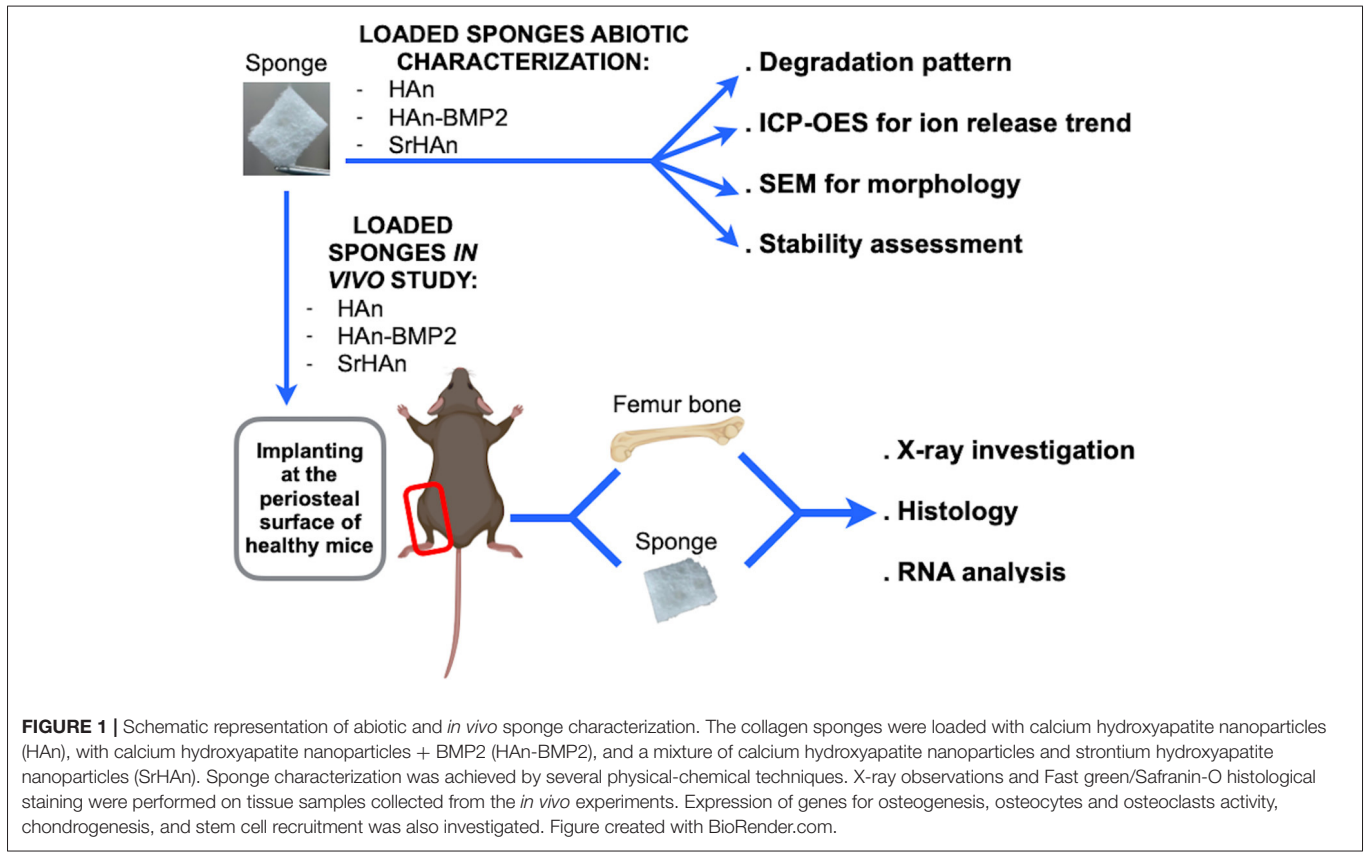

**FIGURE 1** | Schematic representation of abiotic and *in vivo* sponge characterization. The collagen sponges were loaded with calcium hydroxyapatite nanoparticles (HAn), with calcium hydroxyapatite nanoparticles + BMP2 (HAn-BMP2), and a mixture of calcium hydroxyapatite nanoparticles and strontium hydroxyapatite nanoparticles (SrHAn). Sponge characterization was achieved by several physical-chemical techniques. X-ray observations and Fast green/Safranin-O histological staining were performed on tissue samples collected from the *in vivo* experiments. Expression of genes for osteogenesis, osteocytes and osteoclasts activity, chondrogenesis, and stem cell recruitment was also investigated. Figure created with BioRender.com.

A second set of sponges was prepared exactly as the first set of samples but with an additional step of lyophilization in order to check the system stability. Exception made for HAn-BMP2, which was not prepared in the second set. This second set was labeled with an "s" for "stability" (sCTRL, sHAn, sSrHAn) and collectively defined as "treated" sponges. Following the second incubation period carried out in 800 µL of acidified water, all samples were frozen O/N at −20°C and lyophilized for 4 h (−50°C, 0.3 mbar). Following lyophilization samples were re-hydrated with 200 µL of acidified MilliQ. For both sets of sponges, water solutions were poured off and fresh non-acidified MilliQ water was added on day 1, 2, 3, 7, 14, 21, and 28.

### Sponge Calcium, Strontium, and BMP-2 Release

Water solution samples were collected at 1, 2, 3, 7, 14, 21, and 28 days from untreated sponges (CTRL, HAn, SrHAn), treated sponges (sCTRL, sHAn, sSrHAn) and from sponges loaded simultaneously with HAn and rhBMP2 (HAn-BMP2). For the analysis, each sample replicate (1mL) was diluted 1:10 and acidified with suprapur nitric acid (Merck - Sigma Aldrich, Merck KGaA, Darmstadt, Germany) at the final concentration of 0.5% (v/v), centrifuged and filtered with 0.45 µm membranes. Acidified water was used as blank (Moonesi Rad et al., 2019). Measurements were performed using an ICP-OES (inductively coupled plasma optical emission spectroscopy) iCAP 7000 Series (Thermo Scientific) following the standard procedures suggested by the manufacturer and an instrumental method already validated. Calcium and strontium were quantified by external calibration with four standard solutions (0, 5, 1, 5, 10, and 20 mg/L) of the alkaline earth standard metal mix TraceCERT® (100 mg/L). The following wavelengths were selected for the investigated elements: 422,673 nm for $Ca^{2+}$ and 407,771 nm for $Sr^{2+}$. Time zero corresponds to the acidified water before the introduction of the sponges. Statistical evaluation was performed with one-way ANOVA. 95% confidence intervals showing the differences between averaged values were plotted in separated charts, reported as (**Figure S2**). rhBMP2 release rate was evaluated by ELISA. rhBMP2 release in the water solutions was measured after addition of 20 mM monosodium phosphate, in order to displace rhBMP2 from its binding with HAn (Urist et al., 1984; Boix et al., 2005). Samples were then centrifugated to remove excess calcium and 100 µL of each solution was used for immobilization on ELISA wells. Anti-rhBMP2 polyclonal rabbit IgG (ReliaTech GmgH) was used on dilution factor of 1:3,000. Goat anti-rabbit/HRP (Agilent Tech) secondary antibody was used at 1:5,000 dilution. Reaction was developed with TMB (Sigma T0440−100 mL) for 30 min at RT, stopped with sulfuric acid 0.5 M and ELISA microplate was read with Clariostar ELISA reader at 450 nm. (HAn, SrHAn, sHAn and sSrHAn).

### FT-IR Spectroscopy Investigation of Sponge-Nanoparticles Interactions

FT-IR (Fourier Transform-Infra Red) spectra were obtained using a Nicolet FT-IR iS10 Spectrometer (Nicolet, Madison,

WI) equipped with attenuated total reflectance sampling accessory (Smart iTR with diamond plate) by coadding 32 scans in the 4,000–650 $cm^{-1}$ range with a resolution at 4 $cm^{-1}$. FT-IR spectra were recorded on the unloaded sponges, on the unloaded sponges blended together with the nanoparticles (referred to as physical mixture, PM) and on the nanoparticles loaded-sponges.

### Sponge SEM Observations and Mineral Phase Distribution

Untreated (CTRL, HAn, SrHAn) and treated (sCTRL, sHAn, sSrHAn) sponges collected at time 0 days (T0) and time 28 days (T28) were frozen O/N at −20°C and lyophilized for 3 h (−50°C, 0.3 mbar). Microscopic structure was investigated using a scanning electron microscope (SEM) Zeiss EVO-MA10 (Carl Zeiss, Oberkochen, Germany). Images were acquired at different magnifications (150x and 350x) and at an accelerating voltage of 20 kV. CTRL, HAn and SrHAn sponges were also investigated with SEM Zeiss EVO-MA10 (Carl Zeiss, Oberkochen, Germany) coupled to an electron dispersive spectroscopy (EDS) (X-max 50 $mm^2$, Oxford Instruments, Oxford, UK). Images were acquired at a lower magnification and energy dispersive microanalysis was performed in order to map phosphorus (P), calcium (Ca) and strontium (Sr) and quantify carbon (C), oxygen (O), P, ca and Sr.

### Sponge Degradation Rate Measurement

To evaluate the sponge degradation rate, untreated unloaded and loaded sponge masses were measured at the following time points: 1, 2, 3, 7, 14, 21, and 28 days. Sponges were prepared as described above. Day 0 corresponded to the untreated and unloaded dried sponge, while day 1 corresponded to loaded sponges. Weights at day 1 were assumed as the initial weights ($W_i$) and percentage weight loss was calculated for each point. The calculation performed is the following: $[(W_i-W_t)/W_i]*100$, were $W_t$ is the weight at each time point. Average and standard deviation of the percentage weight loss were plotted.

### *In vivo* Animal Study

All animal studies were approved by the Institutional Animal Care and Use Committee at Boston University (BU). Animals enrolled for these studies were crosses of the B6.Cg-Gt(ROSA)26sortm14(CAG-tdTomato)Hze/J (Ai14; stock number 007914) mouse strain with either B6.CG-Pax7tm1(cre/ER2)Gaka/J (Pax7; stock number 017763) or Prx1CreER-GFP (Kawanami et al., 2009). The Ai14 and Pax7 stains were obtained from The Jackson Laboratory (Bar Harbor ME) and housed at the BU animal facilities under standard conditions. All enrolled for this study were healthy, bone damage-free, male mice aged 9–11 weeks. Strains were randomly assigned to treatment and time point.

### Sponge Implant Surgical Procedure and *in vivo* Analysis

The previously prepared untreated samples (HAn, SrHAn, and HAn-BMP2) were used for *in vivo* animal studies. Mice were bilaterally implanted as described before with the exception that the sponges were implanted instead of demineralized bone matrix to induce ectopic bone formation adjacent to the periosteal surface of femurs (Bragdon et al., 2017). Three mice were enrolled per condition per time point (post-operative day 16 and 33) for RNA extraction and histology analysis. **Figure 1** shows a scheme of the performed experiment. No complications and/or collateral effects arose in that period. At post-operative day 16 and 33 mice were euthanized by carbon dioxide inhalation followed by cervical dislocation. Immediately following euthanasia, mice were X-rayed using Faxitron MX-20 Specimen Radiography System at 30 kV for 40 s using Kodak BioMax XAR Scientific Film. The implanted sponge and femur from the left limb were collected separately for RNA extraction and gene analysis. Samples harvested for RNA analysis were stored at −80°C. The right limb was recovered for histology analysis.

### Histological Analysis of Mouse Limb

After fixation in paraformaldehyde (4%), the right limbs were decalcified in 14% w/v EDTA (pH 7.2) for 1 week at 4°C. Limbs were dehydrated and embedded in paraffin for histology and 5 µm-thick sections were cut across the samples. Fast green and Safranin-O (American Mastertek Inc.) staining was performed on the 5 µm-thick sections for the investigation of ectopic bone formation (Provot and Schipani, 2005).

### RNA Extraction and Quantitative Reverse Transcriptase PCR

RNA extraction was performed by tissue dissociation and chemical extraction as previously described (Bragdon et al., 2017). Briefly, samples were snap frozen in QIAzol® Lysis Reagent and lysed with the Qiagen® Tissue Lyser II. Chloroform (Sigma-Aldrich) and isopropanol (Sigma-Aldrich) was used to extract and precipitate RNA followed by 70% ethanol washes. The RNA was re-suspended in 30–50 µL of RNase free water and stored at −80°C. In order to ensure the quality and quantity of the extracted RNA, both spectroscopy and gel electrophoresis were used. For the spectroscopy, a Beckman Coulter™ DU®530 Life Science UV/Vis Spectrophotometer was used and a 260 nm/280 nm ratio value in the range of 1.7–1.9 indicated an acceptable quality of RNA. The 260 nm absorbance value was used to determine the concentration of RNA in the samples. RNA samples were loaded into a 1% agarose gel, GelStar™ Nucleic Acid Gel Stain from Lonza Group was used to detect the presence of the nucleic acid. The presence of two bands under UV light indicates the RNA is intact and not degraded (data not shown). cDNA was synthesized and qRT-PCR was performed as previously described (Bragdon et al., 2017). Briefly, 1 µg of extracted RNA was reverse transcribed using the TaqMan® Reverse Transcription Reagents kit from Applied Biosystems®. Primers used to probe specific expression of genes are listed in **Table 1**. The *18S* gene was used as a reference gene and in addition to our samples non-operative femurs were used as naïve controls. Target gene expression was normalized using the $\Delta\Delta$Ct method of Schmittgen and Livak (2008).

**TABLE 1 |** qRT-PCR gene primers.

| Primer | Catalog number |
| --- | --- |
| **NORMALIZATION PRIMER** | |
| 18S | Mm03928990_gl |
| **STEM CELL RECRUITMENT PRIMER** | |
| Nanog | Mm02384862_gl |
| **CHONDROGENESIS PRIMERS** | |
| Sox9 | Mm00448840_ml |
| Col10A1 | Mm00487041_ml |
| Acan | Mm00545794_ml |
| **OSTEOGENIC- ASSOCIATED PRIMERS** | |
| BGlap | Mm03413826_mH |
| Runx2 | Mm00501578_ml |
| Dmp1 | Mm00803833_gl |
| Sp7 | Mm04209856_sl |
| Ibsp | Mm00492555_ml |
| **OSTEOCLASTS- ASSOCIATED PRIMERS** | |
| Acp5 | Mm00475698_ml |
| Rankl | Mm00441908_ml |
| Ctsk | Mm00484039_ml |
| **OSTEOCYTES- ASSOCIATED PRIMERS** | |
| Sost | Mm00470479_ml |

*List of primers used during the in vivo investigation of genes crucial for the differentiation and activity of cells residing in the bone. The genes are divided over their specificity and their catalog number is reported beside.*

## Statistical Analysis

Statistical analyses were performed with Prism 7 (GraphPad Software Inc). Means are plotted and standard deviations (SD) of means, or standard error means (SEM) are represented by error bars. One-way ANOVA corrected by Tukey's honestly significant difference (THSD) test was used to analyze data, if not otherwise specified. A significance level of 0.05 was used for all statistical analyses, unless noted differently.

## RESULTS

### Physical-Chemical Characterization of the Loaded, Untreated, and Treated Sponges

Untreated sponges (CTRL, HAn, and SrHAn) were prepared and physical-chemically characterized for degradation rate, ion release properties by ICP-OES, sponge-nanoparticles interactions by FT-IR and morphology by SEM. Furthermore, a stability assessment was performed on the treated (lyophilized) sponge samples (sCTRL, sHAn, sSrHAn).

The absolute amounts of calcium and strontium loaded onto each sponge was then calculated. 60 mg of HAn contains 23.89 mg of calcium whereas 54 mg + 6 mg of SrHAn contains 21.5 mg of calcium and 3.55 mg of strontium. Calcium mass for HAn-BMP2 sponge was calculated as for the HAn sample and corresponds to 23.89 mg. ICP-OES data of the water solutions collected at different time points from untreated samples (**Figure 2** and **Table 2**) was used to evaluate the progressive release of calcium and strontium ions in solution.

For HAn-BMP2 samples, calcium ions and rhBMP2 release were measured via ICP-OES and indirect ELISA assay, respectively (**Figures 2C,D** and **Table 2**). An initial burst of $Ca^{2+}$ release was reported for samples HAn and SrHAn (**Figures 2A,B**), and this was followed by a significant decrement at day 3, after which it stabilized at about 10 mg/L—**Table 2**. A similar pattern is presented also for strontium ions (**Figure 2B**), although the concentration values was higher at day 3, ranging between 40 and 50 mg/L—**Table 2**. In Han-BMP2 samples, calcium ion release trend was flattened and strongly decreased (**Figure 2C**) compared to the Han and SrHAn samples. This is probably due to the presence of rhBMP2 in the same solution. Although, the release of rhBMP2 showed a trend very similar to SrHAn samples, with a concentration spike in the first 2 days followed by a decrease at day 3 and constant release until day 28 (**Figure 2D**).

The ICP-OES analyses performed on samples treated for stability assessments showed similar ion release patterns after the lyophilization treatment (**Figure S1** and **Table S1**). In particular, for the sHAn sponges (**Figure S1A**), calcium ion release was higher at day 3 reaching 30 and 40 mg/L. Interestingly, sSrHAn sponges showed a less intense burst at day 3 and increasing values from day 7 to day 28 (**Figure 2D**). Although increments in the days 7–28 were not significant (**Figure S2**). The 95% confidence intervals of the ICP-OES average values are reported in (**Figure S2**).

To evaluate sponge-nanoparticle interactions, FT-IR spectroscopy was employed (**Figure 3**). Unloaded and untreated sponge spectrum (**Figures 3A,B**) showed characteristic protein peaks (Morris and Finney, 2004; Barth, 2007): NH stretching band at 3271 $cm^{-1}$; C=O stretching band (amide I) at 1,624 $cm^{-1}$; N-H bending band (amide II) at 1,528 $cm^{-1}$; C-N stretching (amide III) at 1,232 $cm^{-1}$. FT-IR spectra (**Figures 3C–J**), showed the comparative analysis between unloaded sponge blended along with nanoparticles (physical mixture = PM; **Figures 3C,D,G,H**) and untreated sponges loaded with nanoparticles (**Figures 3E,F,I,J**). All FT-IR spectra of nanoparticles containing-specimens showed good agreement with the characteristic $PO_4^{3-}$ (1,023 $cm^{-1}$) and $OH^-$ (3,570 $cm^{-1}$) vibrations of hydroxyapatite lattice as reported before (Kim et al., 2002). Due to the low concentration of strontium in the samples, no significant variations were recorded between HAn_PM and SrHAn_PM, in accordance with previous results (Frasnelli et al., 2016). However, protein bands showed a much lower intensity in nanoparticles loaded sponges (**Figures 3E,F,I,J**) and completely disappeared in PM spectra (**Figures 3C,D,G,H**). While NH stretching band at 3,271 $cm^{-1}$ was recorded only in the CTRL, C=O stretching and N-H bending slightly moved from initial values (1,624 and 1,528 $cm^{-1}$) to 1,650–1,631 $cm^{-1}$, and 1,531–1,535 $cm^{-1}$, respectively. After lyophilization treatment (**Figure S3**), all protein bands completely disappeared (**Figures S3C–F**).

SEM images were collected to analyze the morphology of unloaded and loaded sponges (**Figure 4**). Unloaded sponges at T0 (**Figure 4A**) were characterized by a spongy structure, with interconnected pores and smooth wall surfaces. Loaded sponge surfaces were rougher and more irregular (**Figures 4B–D**). A reduction of wall surfaces occurred in the unloaded

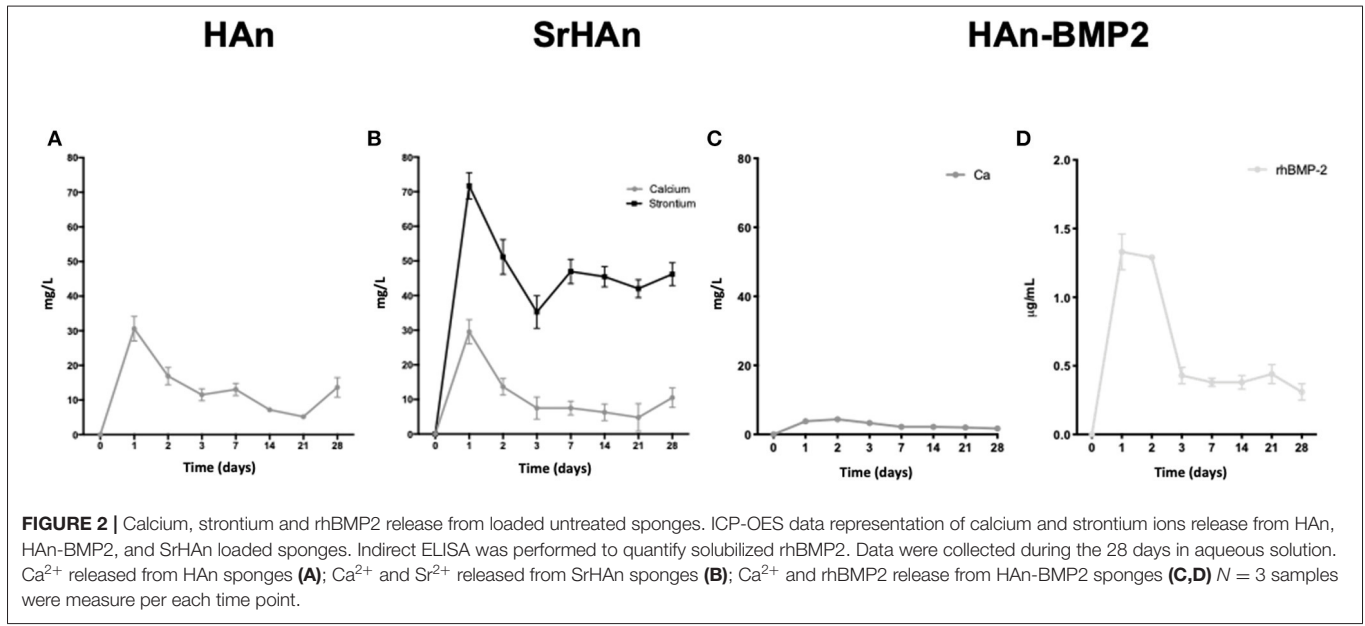

**FIGURE 2** | Calcium, strontium and rhBMP2 release from loaded untreated sponges. ICP-OES data representation of calcium and strontium ions release from HAn, HAn-BMP2, and SrHAn loaded sponges. Indirect ELISA was performed to quantify solubilized rhBMP2. Data were collected during the 28 days in aqueous solution. $Ca^{2+}$ released from HAn sponges **(A)**; $Ca^{2+}$ and $Sr^{2+}$ released from SrHAn sponges **(B)**; $Ca^{2+}$ and rhBMP2 release from HAn-BMP2 sponges **(C,D)** $N = 3$ samples were measure per each time point.

**TABLE 2** | Tabular representation of the data shown in **Figure 2**.

| Time (days) | HAn | SrHAn | | HAn-BMP2 | |
|---|---|---|---|---|---|
| | Ca (mg/L) | Ca (mg/L) | Sr (mg/L) | Ca (mg/L) | rhBMP2 (µg/mL) |
| **0** | 0 | 0 | 0 | 0 | 0 |
| 1 | 30.6 ± 3.5 | 29.5 ± 3.5 | 71.7 ± 3.8 | 3.8 ± 0.4 | 1.3 ± 0.133 |
| 2 | 16.9 ± 2.5 | 13.6 ± 2.4 | 51.1 ± 5.0 | 4.4 ± 0.5 | 1.29 ± 0.01 |
| 3 | 11.5 ± 1.7 | 7.5 ± 3.2 | 35.2 ± 4.7 | 3.3 ± 0.8 | 0.43 ± 0.06 |
| 7 | 13.0 ± 1.8 | 7.5 ± 2.0 | 46.9 ± 3.5 | 2.2 ± 0.4 | 0.38 ± 0.03 |
| 14 | 7.2 ± 0.6 | 6.3 ± 2.4 | 45.4 ± 2.9 | 2.2 ± 0.6 | 0.38 ± 0.03 |
| 21 | 5.2 ± 0.3 | 4.8 ± 4.0 | 42 ± 2.6 | 2.0 ± 0.6 | 0.44 ± 0.07 |
| 2 | 13.7 ± 2.8 | 10.5 ± 2.8 | 46.2 ± 3.3 | 1.7 ± 0.5 | 0.31 ± 0.06 |

*Average concentration values and standard deviations of the calcium and strontium ions and of rhBMP2 measured in the different sponges.*

samples after 28 days (**Figure 4E**), but it was not observed in loaded samples (**Figures 4F–H**). A morphological change was appreciated comparing the T0 and T28 (**Figures 4B,C,E–H**). At T0 hydroxyapatite aggregates were detected on the sponge walls, while at T28 sponge walls looked smoother and the whole structure became more compact. Energy dispersive microanalysis conducted on HAn and SrHAn samples (**Figures 4I–L**) showed a homogenization of the phosphorus and calcium distribution on the sponge surfaces after 28 days in aqueous solution (**Figure 4**, P and Ca distribution maps). The same microanalysis performed on CTRL sample did not show an appreciable level of P, Ca and Sr (data not shown). Moreover, on HAn and SrHAn samples a decrement in the weight percentage of calcium and strontium was recorded at day 28 compared to day 0 (**Figures 4M–P**).

Regarding the treated samples (**Figure S4**), unloaded sponges at T0 (**Figure S4A**) showed a morphology very similar to untreated samples at T0 (**Figure S4A**). Even though, it is possible to appreciate the formation of minor ripples, probably due to the additional lyophilization phase. At T0, the treated and untreated sponges reported similar structures. At T28, treated and loaded sponges acquired a cement-like morphology (**Figures S4F–H**).

The degradation rate was determined by measuring the sponge weight throughout 28 days in aqueous solution (**Figure 5**). A slow degradation pattern was observed with no significant variations recorded, except for the SrHAn loaded sponges (**Figure 5C**). SrHAn samples registered a significant difference in percentage weight loss between day 1 and day 28 ($p < 0.0001$).

A reduction in the quantities of calcium and strontium on the sponge surfaces was recorded at day 28, in accordance with ICP-OES data. Taken together these results indicated the SrHAn loaded sponges are a stable system for a constant and prolonged release of $Ca^{2+}$ and $Sr^{2+}$. They are biodegradable and their morphology fits the golden standard parameters for osteoconductive scaffolds. Although lyophilization partially changed the morphology of the sponges at T28, there was no significant change in calcium and strontium release.

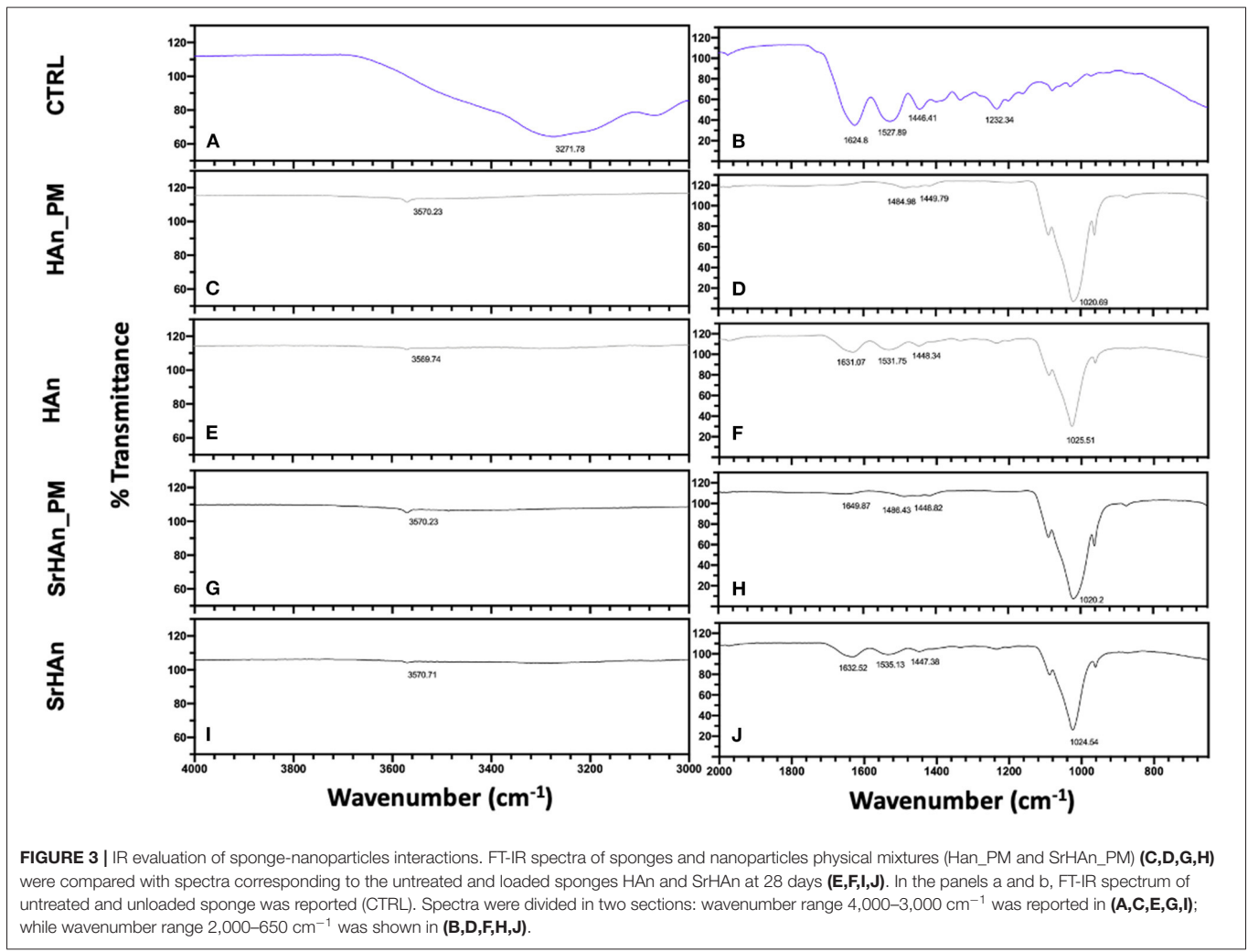

FIGURE 3 | IR evaluation of sponge-nanoparticles interactions. FT-IR spectra of sponges and nanoparticles physical mixtures (Han_PM and SrHAn_PM) (C,D,G,H) were compared with spectra corresponding to the untreated and loaded sponges HAn and SrHAn at 28 days (E,F,I,J). In the panels a and b, FT-IR spectrum of untreated and unloaded sponge was reported (CTRL). Spectra were divided in two sections: wavenumber range 4,000–3,000 cm$^{-1}$ was reported in (A,C,E,G,I); while wavenumber range 2,000–650 cm$^{-1}$ was shown in (B,D,F,H,J).

## *In vivo* Sponge's Implantation and Characterization of Their Effects

Next, sponges prepared following the standard protocol (HAn, HAn-BMP2, and SrHAn) were utilized for *in vivo* implantation to determine the osteogenic ability of the nanoparticles. Radiography, macroscopic observations and histological staining of bone tissues and implanted sponges were carried out to analyze the sponge integration and surrounding tissue adaptation at post-operative day 16 and 33. Gene expression analysis from the implants and femurs were also performed at corresponding times to determine the molecular responses.

The radiographic images support the macroscopic observation that strong integration underwent between the femur, sponge, and surrounding tissue (**Figure 6**). Within the implanted sponges there were few visible blood vessels, as shown in **Figure 7**. Implants and ectopic bones showed various degrees of calcification, as shown by the radiographic images. The sponges loaded with HAn-BMP2 and SrHAn appeared to show increased calcified material than in the sponges with HAn alone. The radiographic and macroscopic images also suggested that there was a reduction in implant dimensions at day 33 compared to day 16 (**Figures 6A–C,J–L**). Since the implants consisted of sponge and hydroxyapatite (black arrowheads in **Figure 6**), it was difficult to distinguish between the implanted material and the formation of ectopic bone. However, the histological analysis using Fast Green and Safranin-O stain demonstrated the formation of both cartilage and bone tissue at the sponge implant site (**Figure 7**).

More extended Safranin-O positive regions (indicated with red arrowheads in **Figure 7**) were observed in the HAn and SrHAn samples at 16 days. Whereas, at 33 days, no Safranin-O positive tissue could be detected. This indicates that the ectopic bone formed through the endochondral ossification process, with cartilaginous tissue turning to calcified areas. The forming bone tissue at the implant sites also appears to support hematopoiesis, as shown by the recruitment of marrow cells between the femur cortical bone and the sponge (**Figure 7E**). The newly formed tissues (bone and bone marrow) were present in all implants but in the SrHAn loaded sponges were more represented. At 33 days, (**Figures 7G–L**) ectopic bone tissue is shown with respect

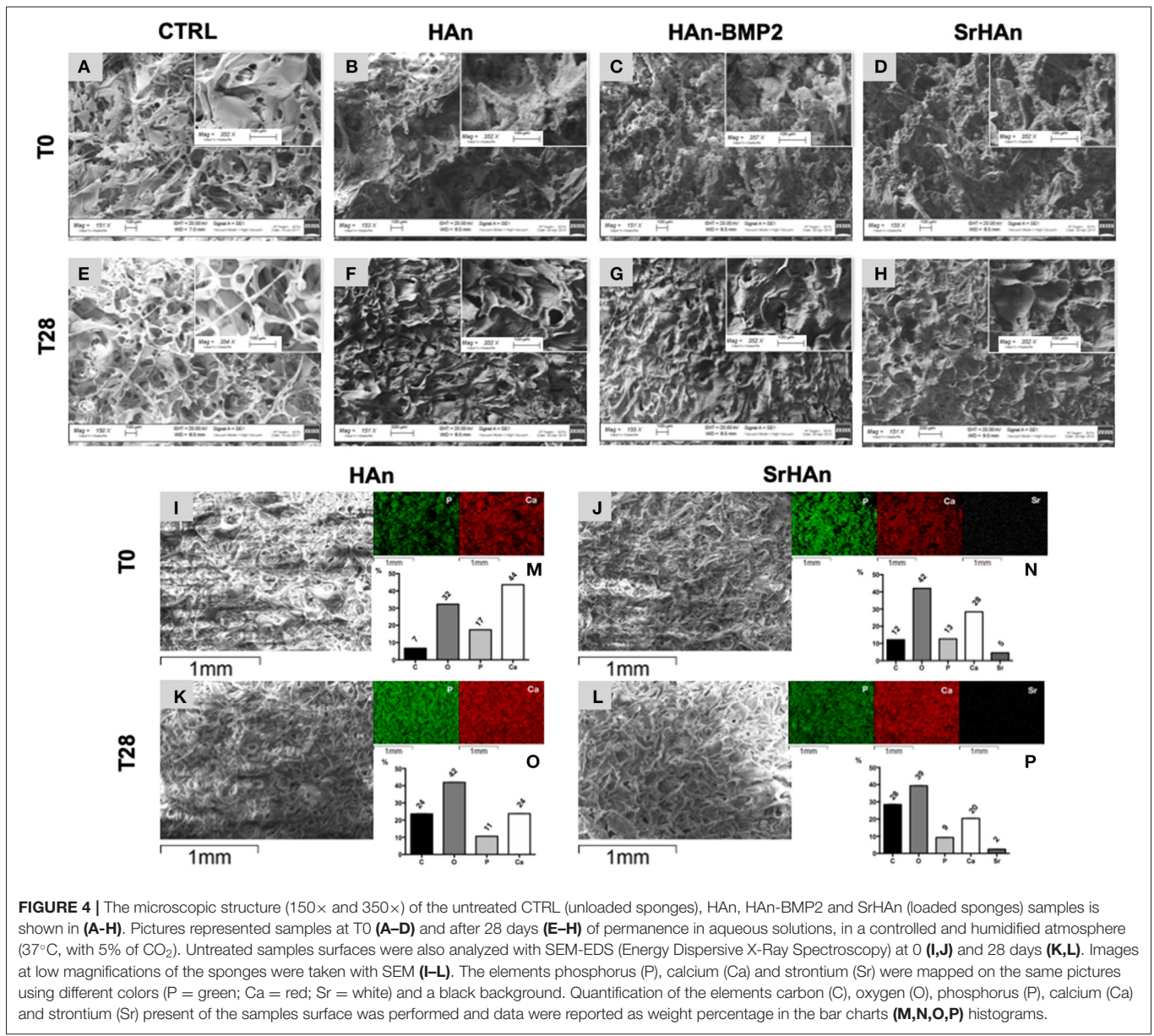

FIGURE 4 | The microscopic structure (150× and 350×) of the untreated CTRL (unloaded sponges), HAn, HAn-BMP2 and SrHAn (loaded sponges) samples is shown in (A–H). Pictures represented samples at T0 (A–D) and after 28 days (E–H) of permanence in aqueous solutions, in a controlled and humidified atmosphere (37°C, with 5% of $CO_2$). Untreated samples surfaces were also analyzed with SEM-EDS (Energy Dispersive X-Ray Spectroscopy) at 0 (I,J) and 28 days (K,L). Images at low magnifications of the sponges were taken with SEM (I–L). The elements phosphorus (P), calcium (Ca) and strontium (Sr) were mapped on the same pictures using different colors (P = green; Ca = red; Sr = white) and a black background. Quantification of the elements carbon (C), oxygen (O), phosphorus (P), calcium (Ca) and strontium (Sr) present of the samples surface was performed and data were reported as weight percentage in the bar charts (M,N,O,P) histograms.

to control, accompanied by the presence of an area enriched in bone marrow cellular components.

To investigate the changes in gene expression in both the femur and implants after 16 and 33 days, tissues were harvested separately from HAn, HAn-BMP2, and SrHAn implanted mice. **Figure 8** represents chondrogenesis (*Acan*, *Col10A1*, and *Sox9*) and stem cell recruitment (*Nanog*) markers. While **Figure 9** shows gene expression of markers for osteogenesis (*Runx2*, *Sp7*, *Ibsp*, *BGlap*, and *Dmp1*), osteocytes (*Sost*) and OCs activity (*Acp5*, *Rankl*, and *Ctsk*). *Acan* (Aggrecan core protein), *Col10A1* (Type 10 collagen) and *Sox9* expression was analyzed as chondrogenesis and endochondral ossification markers (Mackie et al., 2008; Dennis et al., 2015). There was a consistent difference between the expression of chondrogenic markers in bone and sponge samples as noted by the sectioned Y axis. Indeed, at 16 days, in the former these three genes were downregulated by both rhBMP2 and strontium, with respect to HAn condition (**Figure 8A**). While in the latter, both osteoinductive factors brought to significantly higher expression of *Col10A1* and *Sox9* (**Figure 8C**), with an enhanced effect in SrHAn loaded sponges. At 33 days, the difference between the tissue samples is equalized and only significant up-regulation of *Sox9* in the bone and *Acan* in the sponge were shown as strontium-driven effects (**Figures 8E,G**). Expression of the stem cells recruitment marker, *Nanog*, showed no significant difference at 16 days and only a mild up-regulation with respect to control at 33 days in the sponge sample. After 16 days, a significant up-regulation of *BGlap* (Osteocalcin), *Ibsp* (bone sialoprotein), and *Sp7* (Osterix) was observed in the femur of mice that received the SrHAn implant compared to HAn-BMP2 and to HAn samples alone (**Figure 9A**). Similar results were observed with *BGlap, Ibsp,* and *Sp7* expression being evaluated in the sponge samples at 16 days (**Figure 9D**). In both the femur and the sponge tissues *Dmp1* (Dentin matrix protein 1) was upregulated by strontium with respect to the control HAn. A similar trend of expression was also detected at 33 days, where *Ibsp* in the bone (**Figure 9G**) and *Ibsp, BGlap,* and *Sp7* in the sponge (**Figure 9J**) were upregulated by strontium. *Runx2* didn't show significant differences in expression at given time points. *Sost* (Sclerostin) expression was analyzed as marker for osteocyte differentiation. After 16 days, *Sost* was upregulated in the strontium samples only with respect to the control HAn, both in the bone and sponge samples (**Figures 9B,E**). At 33 days, an up-regulation of Sost was observed only in the bone samples loaded with strontium compared to HAn-BMP2 and HAn (**Figures 9H,K**). *Acp5* (Tartrate-resistant acid phosphatase type 5), *Rankl* (Tumor necrosis factor ligand superfamily member 11) and *Ctsk* (Cathepsin K) expression were analyzed as markers for OCs differentiation and activity. At 16 days, *Acp5* and *Rankl* (**Figures 9C,F**) were downregulated in SrHAn femurs and sponges *Ctsk* at 16 days was upregulated in femur and sponge by both types of osteoinductive factors (**Figures 9C,F**). At 33 days, strontium presence led to an up-regulation of *Acp5* and down-regulation of *Rankl* only in the bone samples (**Figure 9I**). *Ctsk* was down-regulated in SrHAn implanted mice with respect to control, both in bone and sponge samples (**Figures 9I,L**).

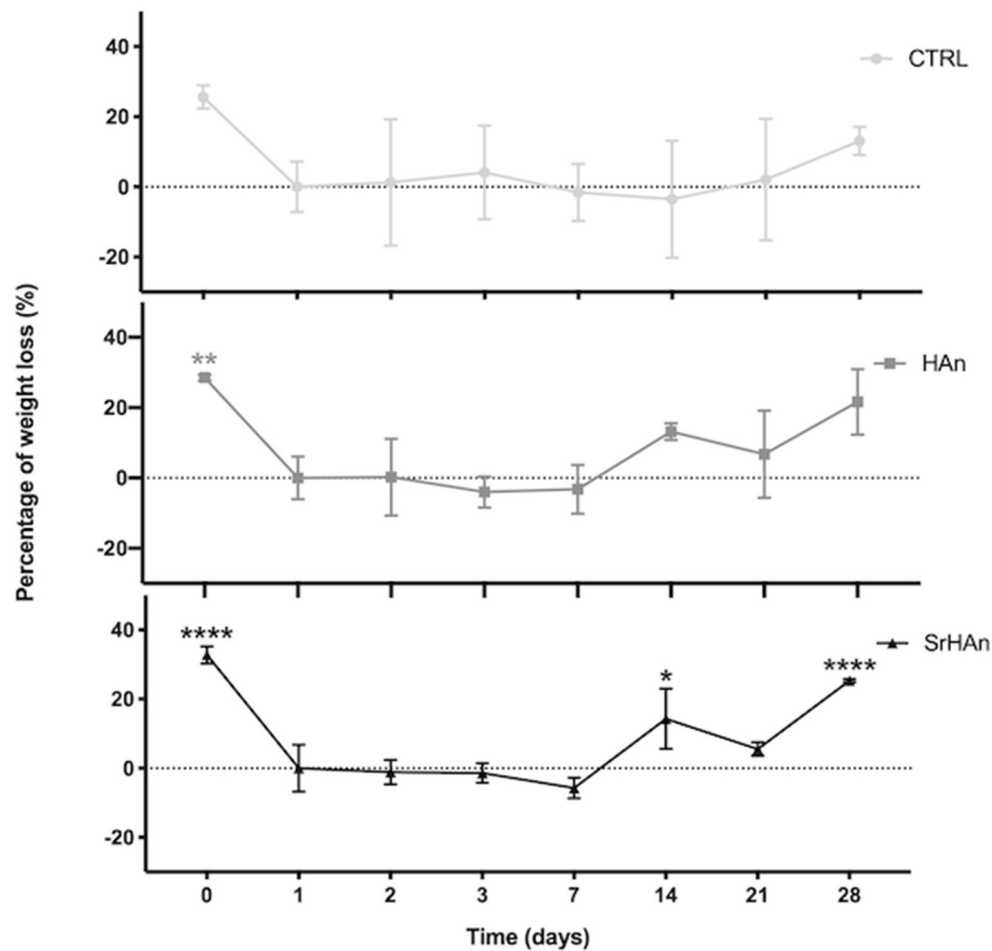

**FIGURE 5** | Graphical representation of sponge weights variations. Untreated sample weights were measured with an analytical balance. Weights were recorded at different time points throughout the 28 days in aqueous solution. Day 0 is referred to the unloaded sponge, whereas Day 1 corresponds to the loaded sponges and it was considered the initial weight (Wi). $N = 3$ different sponges were weighted per each condition. From the recorded weights, we calculated the percentage weight loss. Error bars indicate the standard deviation. Statistically significant differences were assessed with one way ANOVA. Significant differences were reported only against day 1 (*$p < 0,05$; **$p < 0,01$; ****$p < 0,0001$).

SrHAn loaded-sponge structure were suitable for the promotion of cell invasion, bone marrow cell recruitment and ossification. Following a lyophilization treatment (which could be useful for sponge long-term storage and packaging), the loaded sponges did not show significant variation from the untreated samples and all physical-chemical properties were conserved.

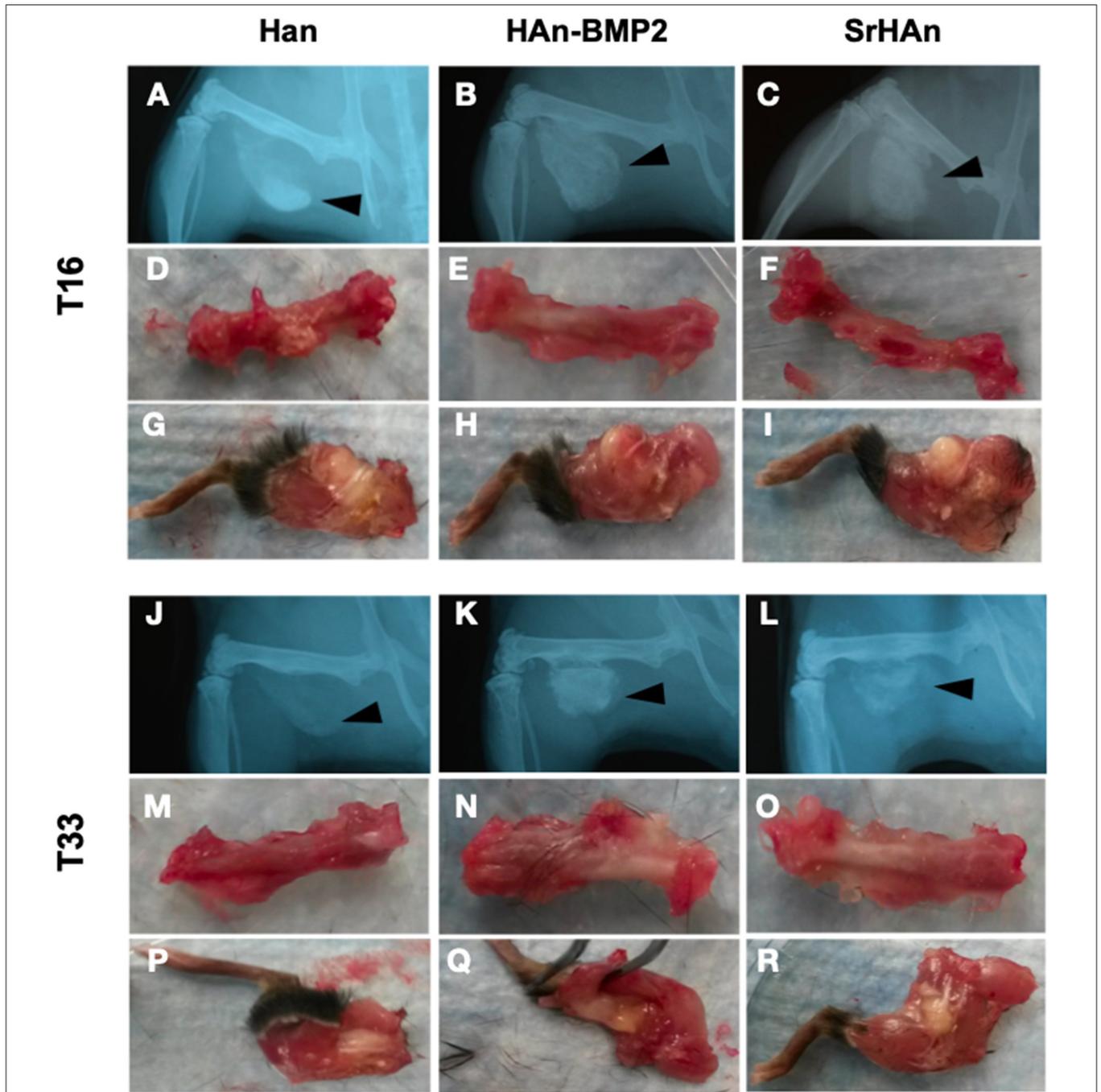

FIGURE 6 | X-ray images and macroscopic observation of mice implanted with loaded sponges. Images at 16 and 33 days of mice implanted with HAn loaded-sponges (A,D,G,J,M,P, respectively), HAn-BMP2 loaded-sponges (B,E,H,K,N,Q) and SrHAn (C,F,I, L,O,R, respectively). The implanted sponges are shown by black arrowheads. Isolated femurs are shown in panels (D–F) and (M–O). Isolated legs including implants are shown in panels (G–I) and (P–R).

The biodegradability and biocompatibility of the system was proven *in vivo*, together with the induction of endochondral ossification stages in adult bone. Evidence of endochondral ossification were given in the histological investigation, where proliferating and hypertrophic chondrocytes were more represented in the strontium-containing samples than in the BMP2 ones. Gene expression corroborated histological evidences, showing increased expression of osteogenic and chondrogenesis markers. Moreover, these data were consistent with an enhancement of endochondral ossification in the SrHAn samples if compared to the HAn-BMP2 samples. Antiresorptive properties of strontium were also demonstrated by investigating markers of osteoclasts differentiation and activity. Histological and gene expression data further

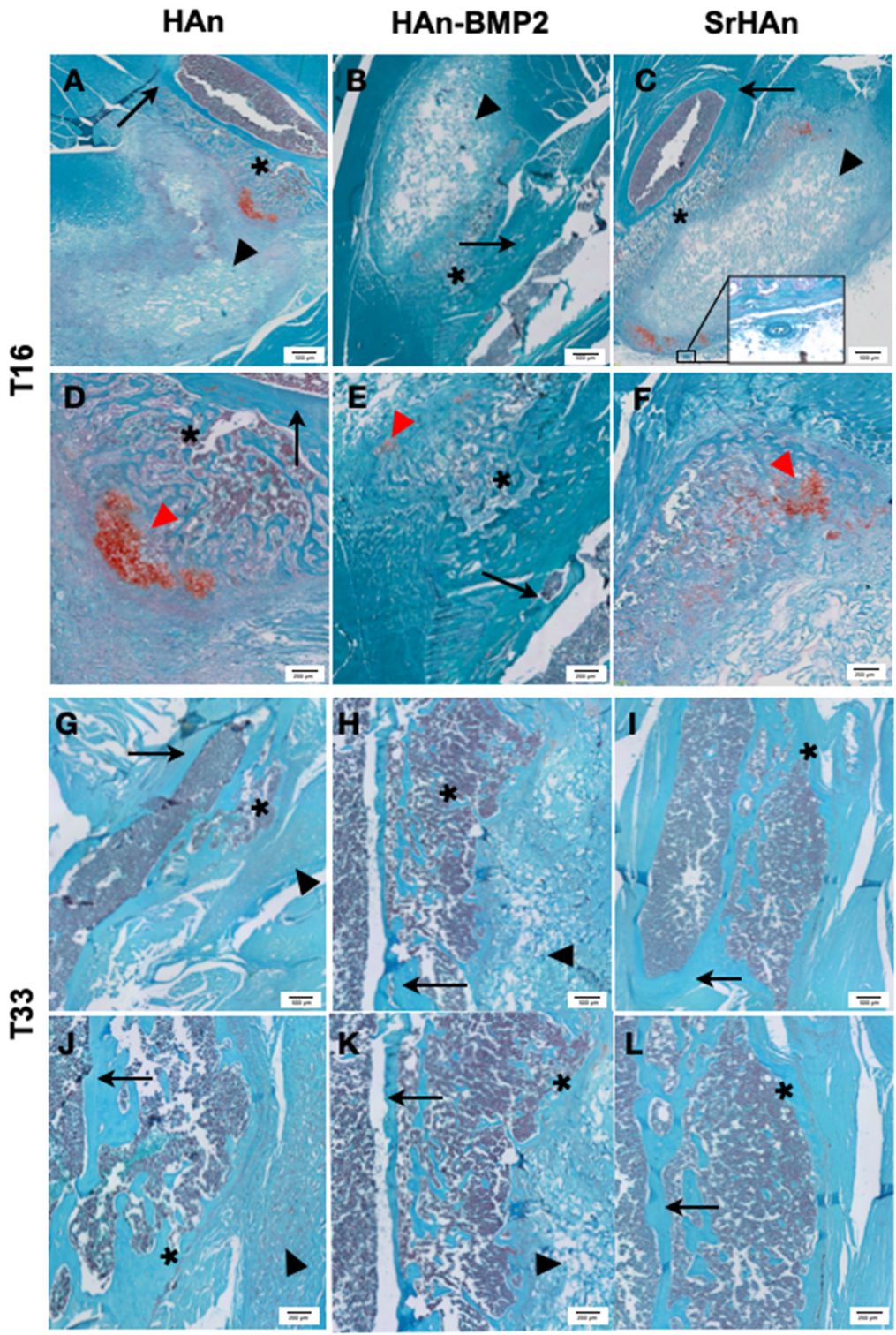

**FIGURE 7** | Representative histological images of post-implants tissues of mouse limb. Fast green/Safranin-O staining was used on post-implant tissue sections of limbs implanted with loaded sponges with HAn **(A,D,G,J)**, HAn-BMP2 **(B,E,H,K)**, or SrHAn **(C,F,I,L)**, for 16 and 33 days, respectively. In the images, red spots of cartilaginous tissue are indicated with *red arrowheads* and gray blurs of ectopic bone are highlighted with *. *Black arrows* and *black arrowheads* indicate femur bone and sponges, respectively. The circle highlights a blood vessel.

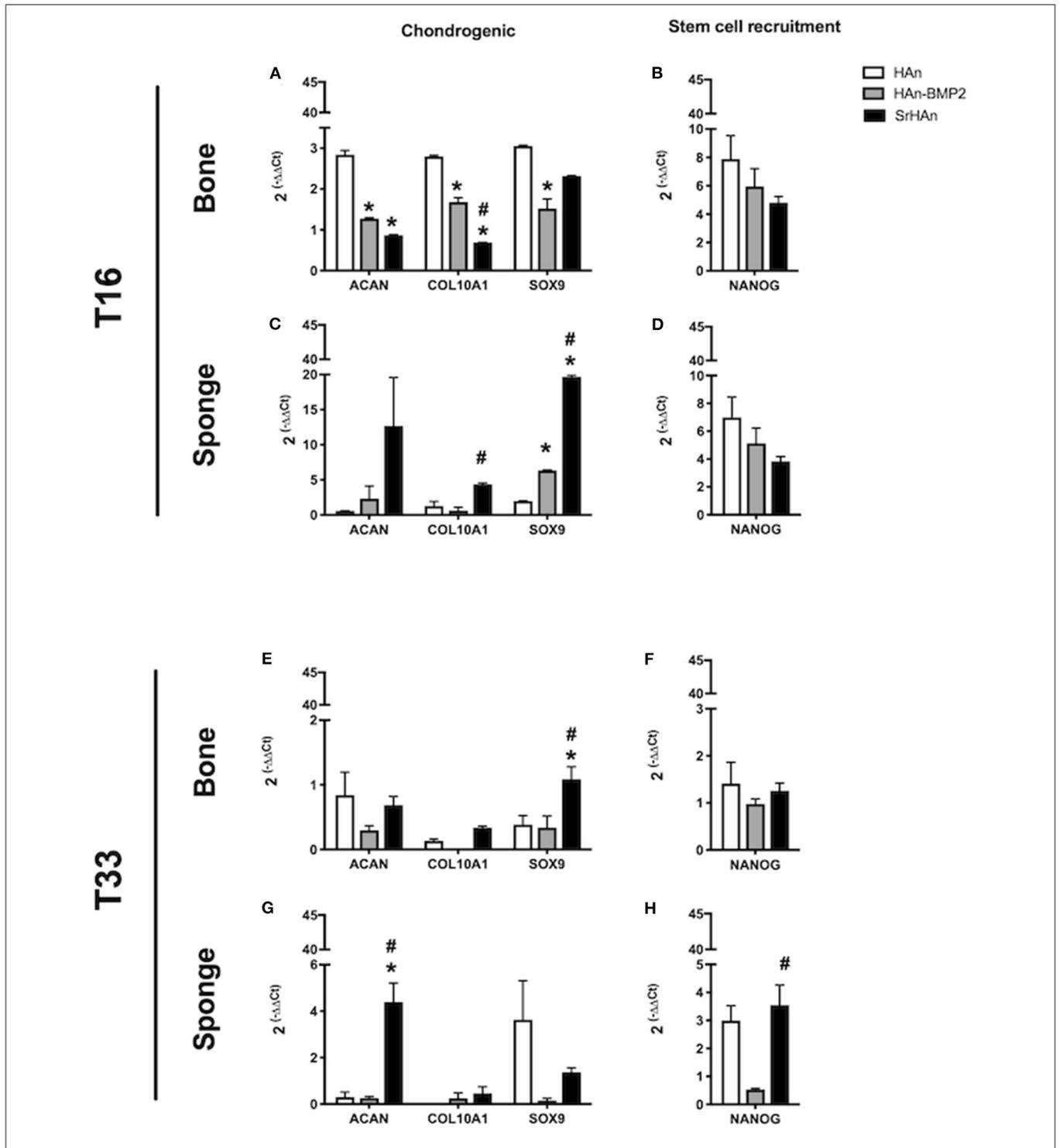

**FIGURE 8** | Gene expression of stem cell recruitment and chondrogenesis markers in bone and sponge post-implants. Gene expression was analyzed in femur bone **(A,B,E,F)** and sponge post-implant **(C,D,G,H)** samples at 16 **(A–D)** and 33 days **(E–H)**, respectively. Expression of chondrogenesis markers Acan, Col10A1 and Sox9 **(A,C,E,G)** as well as stem cell recruitment marker Nanog **(B,D,F,H)** has been evaluated. The graphs show the inverse of the $\Delta\Delta Ct$ at the power of 2. Bars indicate mean values ± SEM of results from 4 experiments. Statistical significance values were calculated with one way-ANOVA, followed by Tukey's honestly significant difference test. *, significant difference against the HAn condition ($p < 0.05$). #, significant difference between HAn-BMP2 and Sr conditions ($p < 0.05$).

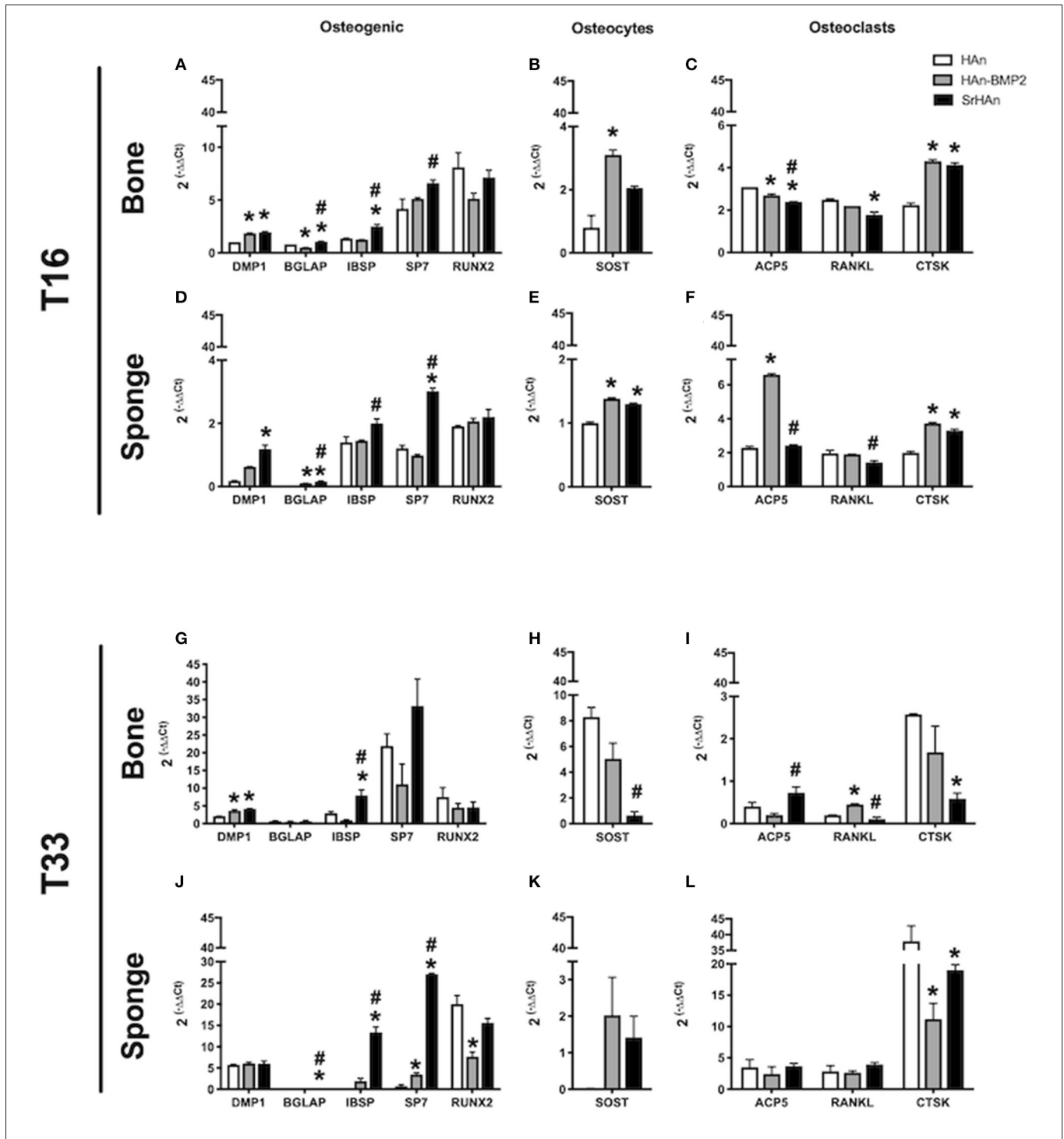

FIGURE 9 | Gene expression of osteocytes, osteoclasts homeostasis, and osteogenic markers in bone and sponge post-implants. Gene expression was evaluated in femur bone (A,B,C,G,H,I) and sponge post-implant (D,E,F,J,K,L) samples at 16 (A–F) and 33 (G–L) days, respectively. Expression of osteogenic markers Dmp1, Bglap, Ibsp, Sp7 and Runx2 (A,D,G,J), osteocytes marker Sost (B,E,H,K) and osteoclasts relevant markers Acp5, Rankl and Ctsk (C,F,I,L) has been evaluated. The graphs show the inverse of the $\Delta\Delta Ct$ at the power of 2. Bars indicate mean values ± SEM of results from four experiments. Statistical significance values were calculated with one way-ANOVA, followed by Tukey's honestly significant difference test. *, significant difference against the HAn condition ($p < 0.05$). #, significant difference between HAn-BMP2 and SrHAn conditions ($p < 0.05$).

supported the osteoinductive and osteoconductive potential of this system.

## DISCUSSION

Bone tissue engineering application in the orthopeadic surgical field is becoming increasingly relevant. In the last decades, rhBMP2 was extensively studied for its osteoinductive potential. Although several successful applications have been achieved, this recombinant morphogen was proven to be not always suitable and safe (Poon et al., 2016; Lykissas and Gkiatas, 2017). Another powerful osteoinductive agent, strontium, was widely studied in the previous years and compelling results were obtained in term of biocompatibility and osteoconductivity (Boanini et al., 2011; Yang et al., 2011). Strontium hydroxyapatite nanoparticles were recently developed, carrying an inorganic osteoinductive agent (Frasnelli et al., 2016). They have been characterized in the last years and in this work, they are presented as a substitute for rhBMP2.

In this study, the medical grade collagen-based Gelfoam sponges were loaded with a combination of either calcium and strontium hydroxyapatite nanoparticles suspensions (SrHAn) or calcium hydroxyapatite nanoparticles and rhBMP2 (HAn-BMP2) in order to compare their physical-chemical properties. The HAn sponges were loaded only with calcium hydroxyapatite nanoparticles and used as control. A second set of samples were treated with an additional lyophilization step and characterized.

The hydroxyapatite nanoparticles amount on each sponge was 30% (w/v) of the volume absorbed by the sponge and 3 µg of rhBMP2 per sponge were used for HAn-BMP2. For the SrHAn loaded-sponges, the calcium/strontium hydroxyapatite nanoparticles ratio was 9:1 w/w (Chandran et al., 2016) while the spontaneously occurring Sr:Ca ratio is between 1:1,000 and 1:2,000 in human bone tissue as well as in blood serum (Cabrera et al., 1999).

The cumulative amount of calcium and strontium ions after 28 days in aqueous solutions were measured as 0.0118 and 0.0358 mg, respectively. Interestingly, several *in vitro* and *in vivo* studies reported that strontium elicits rhBMP2-like effects when administered at concentrations up to 1 mM, as reported by Schumacher et al. (2013) and others (Qiu et al., 2006; Li et al., 2010; Schumacher et al., 2013; Wei et al., 2019). Despite the lower water solubility of $Sr^{2+}$ compared to $Ca^{2+}$, we detected enhanced strontium solubilization from the SrHAn sponges, compared to the calcium release from the HAn sponges. These data are in agreement with Kaufman and Kleinberg (1979) and Pan et al. (2009) studies. They demonstrated how, due to a larger atomic radius, strontium atoms slightly reduce hydroxyapatite lattice thermodynamic stability, favoring the release of $Sr^{2+}$ from the lattice itself. Furthermore, ICP-OES results were comparable to the ones presented by Landi et al. in term of ion release pattern (Landi et al., 2007) and were not significantly influenced by the lyophilization treatment. We investigated the sponge-nanoparticle system through FT-IR analysis of the untreated and treated systems (HAn, SrHAn, sHAn, and sSrHAn). Significant interactions were not found between the sponge and either HAn or SrHAn. FT-IR spectra of loaded sponges showed no chemical interactions were undergoing between the sponges and the nanoparticles. The minor shifting of the amide I and II bands was attributed to hydrogen bonds breaking in the sponge proteins and consequent modification of the hydrogen bonds network in the sponge structure. We can therefore assume that the ion release patterns studied with ICP-OES were not influenced by chemical interactions between the nanoparticles and the sponge support, either before or after the lyophilization treatment. To illustrate the patterns of ion release in function of time we considered the porous structure of the sponges. We could speculate a bulk of nanoparticles flowed inside the sponge pores, during the loading procedure. Nanoparticles remaining on the sponge surfaces released a high percentage of $Ca^{2+}$ and $Sr^{2+}$ ions during the first 3 days, while embedded nanoparticles were responsible for the constant release recorded after day 3. The same speculation can also fit the data on rhBMP2 release. This hypothesis was corroborated by SEM images, that showed smoother surfaces and solubilization of hydroxyapatite aggregates at day 28. Surprisingly, even after 28 days in aqueous solution, calcium and traces of strontium were still detected on the sponge surfaces, as demonstrated by SEM-EDS data.

Interestingly, the calcium release pattern of HAn-BMP2 loaded-sponges was different from the other samples. This can be due to the non-covalent interaction between HAn and rhBMP2 (Urist et al., 1984; Boix et al., 2005). We were not able to investigate HAn-BMP2 loaded-sponges with FT-IR, since this technique is not able to discern between different proteins. We can speculate that calcium ions remained bound to solubilized rhBMP2, therefore decreasing the amount of free $Ca^{2+}$ released in solution.

Pore sizes ranging between 100 and 300 µm are considered the most suitable for osteoinduction improvement as well as for angiogenesis promotion (Karageorgiou and Kaplan, 2005; Scheinpflug et al., 2018). Furthermore, geometry of the investigated sponges reflected the "gold standard" in bone tissue engineering, according to Kuboki et al. (2007) and due to the given pores size and geometry the surface area exposed to the surrounding microenvironment is increased, allowing for a wider interaction between the nanoparticles coating and the surrounding microenvironment.

The degradability and recyclability of the graft is a crucial point, although often disregarded, in tissue engineering applications (Gentleman et al., 2010; Rohnke et al., 2016). In this work, the time-dependent biodegradability of the sponges and the hydroxyapatite nanoparticles both in the abiotic and *in vivo* conditions, were analyzed. Degradation pattern based on the sponge weights appeared mild but significant for SrHAn sponges, throughout the 28 days in solution. While, *in vivo* radiographic images and histology suggested an accelerated degradation of SrHAn sponges. Radiographic images also revealed an induction of bone formation and osseointegration by all three types of implants at both time points. Although, a more extensive mineralized tissue in mice implanted with HAn-BMP2 and SrHAn sponges were shown with respect to the controls. Histological analysis revealed new tissue formation between the sponge and the femoral cortex, suggesting a center of active osteogenesis was present in the sponge site. Given the similarities between $Ca^{2+}$ and $Sr^{2+}$, it has been shown that this element can substitute calcium in the lattice of the hydroxyapatite crystals present in bones (Ni et al., 2011; Querido et al., 2016). $Sr^{2+}$ is processed like $Ca^{2+}$ throughout metabolism, being preferentially introduced in active osteogenesis sites (Bauman et al., 2005).

RhBMP2 and $Sr^{2+}$ release was quantified using different techniques, therefore it is not possible to compare their quantities/concentrations in solution head-to-head. Despite the similarities in the pattern release between HAn-BMP2 and SrHAn, strontium containing nanoparticles induced significant increase of endochondral ossification markers. It is well-known how the process of endochondral ossification can be recapitulated in adults by bone fracture and bone tissue damage (Gerstenfeld et al., 2003; Dennis et al., 2015), although in this work evidences of endochondral ossification were induced in healthy implanted mice, with no bone damage. Furthermore, no record of strontium related-endochondral ossification induction is present up to date, but here at 16 days the onset of typical multicellular clusters constitutes by enlarged chondrocyte was demonstrated (Mackie et al., 2008). Hypertrophic chondrocytes can be detected in the sponge surroundings at 16 days, together with smaller proliferating chondrocytes. These were particularly represented in the HAn and SrHAn samples and red stained-cartilage proteoglycans were present all around the cells. Hypertrophic chondrocytes were almost absent in the HAn-BMP2 samples, suggesting a possible delay or acceleration in chondrogenic differentiation. Different phases of endochondral ossification process were shown also by gene expression data: *Sox9* upregulation was an indicator of chondrocytes proliferation (Akiyama et al., 2002; Lefebvre and Dvir-Ginzberg, 2017) and after its downregulation, markers of hypertrophic chondrocytes, such as aggrecan core protein (*Acan*) and type X collagen (*Col10A1*) were upregulated (Akiyama and Lefebvre, 2011). We can speculate that the different temporal upregulation of *Sox9*, among bone and sponges, can be attributed to the time dependent $Sr^{2+}$ diffusion, which resulted in its earlier expression in the sponge area and delayed expression in the bone one.

The interaction of $Sr^{2+}$ with the calcium-sensing receptor (CaSR) was shown both in OBs and osteoclasts (OCs), resulting in improved anabolic processes and reduced catabolic pathways, respectively (Chattopadhyay et al., 2006). Our gene expression data showed SrHAn double activity on OBs and OCs. Several studies reported that strontium modulates alkaline phosphatase (ALP) activity and Runx-2 expression (Tian et al., 2014) in OBs and OBs progenitors while promoting Wnt/β-catenin pathway in human bone marrow derived mesenchymal stem cells (MSCs), thereby enhancing ECM secretion and osteogenic differentiation (Yang et al., 2011). Interestingly, in our study no significant variation was detected in *Runx2* levels. However, our results measured an upregulation of *Sp7* (Osterix), indicating OBs differentiation at early and late stage (Santo et al., 2013; Kawane et al., 2018). Sp7 is a transcriptional factor expressed in developing bones regulating the commitment of MSCs toward osteoblastic lineage (Rutkovskiy et al., 2016). Its regulation affects expression of protein-coding genes typically related to osteoblastic differentiation, such as osteocalcin, type I collagen and bone sialoprotein (*BGlap, Ibsp,* and *Dmp1*). Increased expression of the ECM proteins BGlap, Ibsp, and Dmp1 was attributed to mature osteoblasts activity (Wrobel et al., 2016) in SrHAn samples. Sp7, BGlap, and Ibsp expression also indicated a stronger osteogenic induction for strontium ions compared to rhBMP2. Interestingly Runx2, the master regulator of osteoblastic differentiation acting above Sp7, was not significantly influenced neither by strontium nor by rhBMP2 (Nakashima et al., 2002). Soluble cytokine promoting OCs differentiation, Rankl, (expressed by OBs) is a key modulator of the resorption cycle, controlling OCs differentiation and activation, and therefore bone resorption. This molecule has already been linked to strontium-mediated modulation by Atkins et al. (2009) and Brennan et al. (2009). Our results showed strontium significantly downregulated *Rankl* when compared to control and HAn-BMP2 samples. It has been shown that strontium effects are mediated by CaSR and by others, yet unclarified receptors (Chattopadhyay et al., 2006; Takaoka et al., 2010; Saidak and Marie, 2012). Furthermore, being CaSR present also on OCs outer membrane it was demonstrated the action of strontium on OCs activity. Indeed, OCs-secreted endoprotease responsible for ECM degradation (Sage et al., 2012), Ctsk (Cathepsin K) was reduced by strontium presence at 33 days, with respect to the control. Of note, in some of the SrHAn sponges we observed vascularization, suggesting a potential effect of $Sr^{2+}$ on neo-vascularization, although further studies are needed to confirm these findings.

## CONCLUSIONS

Many studies have been conducted to investigate rhBMP2 suitability as treatment for spinal fractures and critically-sized bone defects. Although its indisputable osteoinductive potential, rhBMP2 also showed adverse effects. On the other hand, strontium has already been tested as a treatment for osteoporosis and its activity was demonstrated in a patient population. The biggest advantage of strontium over rhBMP2 however, is that the former is a chemical and not a biologic. Herein, following the abiotic characterization, we presented the results of a comparative *in vivo* study, where Gelfoam sponges loaded either with strontium hydroxyapatite nanoparticles or with rhBMP2 were implanted in healthy, bone damage-free mouse model. *In vivo* studies, showed that SrHAn and HAn-BMP2 have comparable effects, driving the onset of endochondral ossification and promoting the bone remodeling process. These results demonstrated that SrHAn loaded-sponges have marked osteogenic potential when applied on the periosteum of long bones, comparable to HAn-BMP2 loaded-sponges, but eliciting a more controlled ossification response. We propose to use Gelfoam sponges enriched in $Sr^{2+}$ as an effective therapeutic intervention to treat severe bone defects or open, complicated fractures. The short-term benefit of this nanotechnological system is the availability of novel therapeutic option to treat bone fracture (or bone defect). The use of sponges loaded with strontium hydroxyapatite nanoparticles might also provide better outcomes for complex fractures. Results from these studies can provide novel therapeutic options for active duty personnel and can be beneficial to anyone suffering from trauma, bone defects or severe bone injuries. Future studies shall also evaluate the efficacy of SrHAn loaded sponges for the treatment of spinal fusion.

## DATA AVAILABILITY STATEMENT

All datasets generated for this study are included in the article/**Supplementary Material**.

## ETHICS STATEMENT

All animal studies were approved by the Institutional Animal Care and Use Committee at Boston University (BU).

## AUTHOR CONTRIBUTIONS

GM and FC: conceptualization, validation, formal analysis, investigation, writing—original draft and review and editing, and vaisualization. GB and LC: validation, formal analysis, investigation, resources, writing—review & editing, and visualization. LF and AK: conceptualization, validation, formal analysis, writing—review & editing. PD: conceptualization, formal analysis, resources, writing—original draft & review & editing, visualization, supervision, project administration, and funding acquisition. BB: validation, formal analysis, investigation, writing—original draft & review & editing, visualization, supervision, and project administration. LV and LG: conceptualization, resources, writing—original draft & review & editing, visualization, supervision, project administration, and funding acquisition.

## FUNDING

This research was supported by funds from the National Institutes of Health NIAMS K99AR068582 (BB), NIAMS R01AR056637 (LG), and a grant from the Musculoskeletal Transplant Foundation (BB). The content of this study is solely the responsibility of the authors and do not necessarily represent the official views of the NIH. This work also was supported by grants from the Italian Space Agency, Project DC-MIC-2012-024, and contract N. 2013-060-I.O (to LV) and Projects of High Relevance, Bilateral project Italy-Sweden, Ministry of Foreign Affairs and International Cooperation (MAECI), and Ministry of Education, University and Research (MIUR) (2018-2020) titled "Effect of Microgravity e nanoparticles on bone regeneration in simulated Microgravity (REPAIR)" (to LV). Furthermore, a grant of the Italian Ministry of Education, University and Research (MIUR) to the Department of Molecular Medicine of the University of Pavia under the initiative "Dipartimenti di Eccellenza (2018–2022)" also contributed to this research.

## ACKNOWLEDGMENTS

We would like to acknowledge Anne Hinds for technical assistance with sectioning histological samples. We would like to thank professor Paola Petrini (Dipartimento di Chimica, Materiali ed Ingegneria Chimica G. Natta, Politecnico Milano, Italy) for the theoretical contribution on the formulation of the nanoparticles loaded-sponges.

## SUPPLEMENTARY MATERIAL

The Supplementary Material for this article can be found online at: https://www.frontiersin.org/articles/10.3389/fbioe.2020.00499/full#supplementary-material

## REFERENCES

Akiyama, H., Chaboissier, M. C., Martin, J. F., Schedl, A., and De Crombrugghe, B. (2002). The transcription factor Sox9 has essential roles in successive steps of the chondrocyte differentiation pathway and is required for expression of Sox5 and Sox6. *Genes Dev*. 16, 2813–2828. doi: 10.1101/gad.1017802

Akiyama, H., and Lefebvre, V. (2011). Unraveling the transcriptional regulatory machinery in chondrogenesis. *J. Bone Miner. Metab*. 29, 390–395. doi: 10.1007/s00774-011-0273-9

Atkins, G. J., Welldon, K. J., Halbout, P., and Findlay, D. M. (2009). Strontium ranelate treatment of human primary osteoblasts promotes an osteocyte-like phenotype while eliciting an osteoprotegerin response. *Osteoporos. Int*. 20, 653–664. doi: 10.1007/s00198-008-0728-6

Barth, A. (2007). Infrared spectroscopy of proteins. *Biochim. Biophys. Acta Bioenerg*. 1767, 1073–1101. doi: 10.1016/j.bbabio.2007.06.004

Bauman, G., Charette, M., Reid, R., and Sathya, J. (2005). Radiopharmaceuticals for the palliation of painful bone metastases - a systematic review. *Radiother. Oncol*. 75, 258–70. doi: 10.1016/j.radonc.2005.03.003

Boanini, E., Torricelli, P., Fini, M., and Bigi, A. (2011). Osteopenic bone cell response to strontium-substituted hydroxyapatite. *J. Mater. Sci. Mater. Med*. 22, 2079–2088. doi: 10.1007/s10856-011-4379-3

Boix, T., Gómez-Morales, J., Torrent-Burgués, J., Monfort, A., Puigdomènech, P., and Rodríguez-Clemente, R. (2005). Adsorption of recombinant human bone morphogenetic protein rhBMP-2m onto hydroxyapatite. *J. Inorg. Biochem*. 99, 1043–1050. doi: 10.1016/j.jinorgbio.2005.01.011

Boraiah, S., Paul, O., Hawkes, D., Wickham, M., and Lorich, D. G. (2009). Complications of recombinant human BMP-2 for treating complex tibial plateau fractures: a preliminary report. *Clin. Orthop. Relat. Res*. 467, 3257–3262. doi: 10.1007/s11999-009-1039-8

Bragdon, B., Lam, S., Aly, S., Femia, A., Clark, A., Hussein, A., et al. (2017). Earliest phases of chondrogenesis are dependent upon angiogenesis during ectopic bone formation in mice. *Bone* 101, 49–61. doi: 10.1016/j.bone.2017.04.002

Brennan, T. C., Rybchyn, M. S., Green, W., Atwa, S., Conigrave, A. D., and Mason, R. S. (2009). Osteoblasts play key roles in the mechanisms of action of strontium ranelate. *Br. J. Pharmacol*. 157, 1291–300. doi: 10.1111/j.1476-5381.2009.00305.x

Buza, J. A., and Einhorn, T. (2016). Bone healing in 2016. *Clin. Cases Miner. Bone Metab*. 13, 101–105. doi: 10.11138/ccmbm/2016.13.2.101

Cabrera, W. E., Schrooten, I., De Broe, M. E., and D'Haese, P. C. (1999). Strontium and bone. *J. Bone Miner. Res*. 14, 661–668. doi: 10.1359/jbmr.1999.14.5.661

Carmo, A. B. X. D., Sartoretto, S. C., Alves, A. T. N. N., Granjeiro, J. M., Miguel, F. B., Calasans-Mala, J., et al. (2018). Alveolar bone repair with strontiumcontaining nanostructured carbonated hydroxyapatite. *J. Appl. Oral Sci*. 26:e20170084. doi: 10.1590/1678-7757-2017-0084

Chandran, S., Shenoy, S. J., Babu, S. S., Nair, P. R., H. K. V., and John, A. (2018). Strontium hydroxyapatite scaffolds engineered with stem cells aid osteointegration and osteogenesis in osteoporotic sheep model. *Colloids Surf. B Biointerfaces* 163, 346–354. doi: 10.1016/j.colsurfb.2017.12.048

Chandran, S., Suresh Babu, S., Hari Krishnan, V. S., Varma, H. K., and John, A. (2016). Osteogenic efficacy of strontium hydroxyapatite microgranules in osteoporotic rat model. *J. Biomater. Appl.* 31, 499–509. doi: 10.1177/0885328216647197

Chattopadhyay, N., Brown, E. M., Terwilliger, E. F., Petit, L., Yano, S., Brazier, M., et al. (2006). The calcium sensing receptor is directly involved in both osteoclast differentiation and apoptosis. *FASEB J.* 20, 2562–2564. doi: 10.1096/fj.06-6304fje

Cristofaro, F., Pani, G., Pascucci, B., Mariani, A., Balsamo, M., Donati, A., et al. (2019). The NATO project: nanoparticle-based countermeasures for microgravity-induced osteoporosis. *Sci. Rep.* 9:17141. doi: 10.1038/s41598-019-53481-y

Dennis, S. C., Berkland, C. J., Bonewald, L. F., and Detamore, M. S. (2015). Endochondral ossification for enhancing bone regeneration: converging native extracellular matrix biomaterials and developmental engineering *in vivo*. *Tissue Eng. B Rev.* 21, 247–266. doi: 10.1089/ten.teb.2014.0419

Frasnelli, M., Cristofaro, F., Sglavo, V. M., Dirè, S., Callone, E., Ceccato, R., et al. (2016). Synthesis and characterization of strontium-substituted hydroxyapatite nanoparticles for bone regeneration. *Mater. Sci. Eng. C* 71, 653–662. doi: 10.1016/j.msec.2016.10.047

Friess, W., Uludag, H., Foskett, S., Biron, R., and Sargeant, C. (1999). Characterization of absorbable collagen sponges as rhBMP-2 carriers. *Int. J. Pharm.* 187, 91–99. doi: 10.1016/S0378-5173(99)00174-X

Gentleman, E., Fredholm, Y. C., Jell, G., Lotfibakhshaiesh, N., O'Donnell, M. D., Hill, R. G., et al. (2010). The effects of strontium-substituted bioactive glasses on osteoblasts and osteoclasts *in vitro*. *Biomaterials* 31, 3949–3956. doi: 10.1016/j.biomaterials.2010.01.121

Gerstenfeld, L. C., Cullinane, D. M., Barnes, G. L., Graves, D. T., and Einhorn, T. A. (2003). Fracture healing as a post-natal developmental process: molecular, spatial, and temporal aspects of its regulation. *J. Cell. Biochem.* 88, 873–884. doi: 10.1002/jcb.10435

Giorgi, P., Capitani, D., Sprio, S., Sandri, M., Tampieri, A., Canella, V., et al. (2017). A new bioinspired collagen-hydroxyapatite bone graft substitute in adult scoliosis surgery: results at 3-year follow-up. *J. Appl. Biomater. Funct. Mater.* 15, e262–e270. doi: 10.5301/jabfm.5000366

Hoffmann, M. F., Jones, C. B., and Sietsema, D. L. (2013). Complications of rhBMP-2 utilization for posterolateral lumbar fusions requiring reoperation: a single practice, retrospective case series report. *Spine J.* 13, 1244–52. doi: 10.1016/j.spinee.2013.06.022

Holroyd, C., Cooper, C., and Dennison, E. (2008). Epidemiology of osteoporosis. *Best Pract. Res. Clin. Endocrinol. Metab.* 22, 671–85. doi: 10.1007/978-3-211-74221-1_1

Karageorgiou, V., and Kaplan, D. (2005). Porosity of 3D biomaterial scaffolds and osteogenesis. *Biomaterials* 26, 5474–91. doi: 10.1016/j.biomaterials.2005.02.002

Kaufman, H. W., and Kleinberg, I. (1979). Studies on the incongruent solubility of hydroxyapatite. *Calcif. Tissue Int.* 27, 143–151. doi: 10.1007/BF02441177

Kawanami, A., Matsushita, T., Chan, Y. Y., and Murakami, S. (2009). Mice expressing GFP and CreER in osteochondro progenitor cells in the periosteum. *Biochem. Biophys. Res. Commun.* 386, 477–82. doi: 10.1016/j.bbrc.2009.06.059

Kawane, T., Qin, X., Jiang, Q., Miyazaki, T., Komori, H., Yoshida, C. A., et al. (2018). Runx2 is required for the proliferation of osteoblast progenitors and induces proliferation by regulating Fgfr2 and Fgfr3. *Sci. Rep.* 8:13551. doi: 10.1038/s41598-018-31853-0

Kim, J. H., Kim, S. H., Kim, H. K., Akaike, T., and Kim, S. C. (2002). Synthesis and characterization of hydroxyapatite crystals: a review study on the analytical methods. *J. Biomed. Mater. Res.* 62, 600–612. doi: 10.1002/jbm.10280

Kowalczewski, C. J., and Saul, J. M. (2018). Biomaterials for the delivery of growth factors and other therapeutic agents in tissue engineering approaches to bone regeneration. *Front. Pharmacol.* 9:513. doi: 10.3389/fphar.2018.00513

Kuboki, Y., Jin, Q., Kikuchi, M., Mamood, J., and Takita, H. (2007). Geometry of artificial ECM: sizes of pores controlling phenotype expression in BMP-induced osteogenesis and chondrogenesis. *Connect. Tissue Res.* 43, 529–534. doi: 10.1080/03008200290001104

Landi, E., Tampieri, A., Celotti, G., Sprio, S., Sandri, M., and Logroscino, G. (2007). Sr-substituted hydroxyapatites for osteoporotic bone replacement. *Acta Biomater.* 3, 961–969. doi: 10.1016/j.actbio.2007.05.006

Lanza, D., and Vegetti, M. (1974). *Opere Biologiche*. Aristotele, Diego Lanza, Mario Vegetti: Isis.

Latzman, J. M., Kong, L., Liu, C., and Samadani, U. (2010). Administration of human recombinant bone morphogenetic protein-2 for spine fusion may be associated with transient postoperative renal insufficiency. *Spine (Phila. Pa. 1976).* 35, E231–7. doi: 10.1097/BRS.0b013e3181c71447

Lefebvre, V., and Dvir-Ginzberg, M. (2017). SOX9 and the many facets of its regulation in the chondrocyte lineage. *Connect. Tissue Res.* 58, 2–14. doi: 10.1080/03008207.2016.1183667

Li, J., Yang, L., Guo, X., Cui, W., Yang, S., Wang, J., et al. (2018). Osteogenesis effects of strontium-substituted hydroxyapatite coatings on true bone ceramic surfaces *in vitro* and *in vivo*. *Biomed. Mater.* 13:015018. doi: 10.1088/1748-605X/aa89af

Li, Y., Li, Q., Zhu, S., Luo, E., Li, J., Feng, G., et al. (2010). The effect of strontium-substituted hydroxyapatite coating on implant fixation in ovariectomized rats. *Biomaterials* 31, 9006–14. doi: 10.1016/j.biomaterials.2010.07.112

Luca, L., Capelle, M. A. H., Machaidze, G., Arvinte, T., Jordan, O., and Gurny, R. (2010). Physical instability, aggregation and conformational changes of recombinant human bone morphogenetic protein-2 (rhBMP-2). *Int. J. Pharm.* 391, 48–54. doi: 10.1016/j.ijpharm.2010.02.015

Lykissas, M., and Gkiatas, I. (2017). Use of recombinant human bone morphogenetic protein-2 in spine surgery. *World J. Orthop.* 8, 531–535. doi: 10.5312/wjo.v8.i7.531

Mackie, E. J., Ahmed, Y. A., Tatarczuch, L., Chen, K. S., and Mirams, M. (2008). Endochondral ossification: how cartilage is converted into bone in the developing skeleton. *Int. J. Biochem. Cell Biol.* 40, 46–62. doi: 10.1016/j.biocel.2007.06.009

Marcus, R. (2007). *Fundamentals of Osteoporosis. Third*, eds D. Feldman, D. A. Nelson, and C. J. (Rosen Burlington, MA: Elsevier Inc.).

Moonesi Rad, R., Pazarçeviren, E., Ece Akgün, E., Evis, Z., Keskin, D., Sahin, S., et al. (2019). *In vitro* performance of a nanobiocomposite scaffold containing boron-modified bioactive glass nanoparticles for dentin regeneration. *J. Biomater. Appl.* 33, 834–853. doi: 10.1177/0885328218812487

Morris, M. D., and Finney, W. F. (2004). Recent developments in Raman and infrared spectroscopy and imaging of bone tissue. *Spectroscopy* 18, 155–159. doi: 10.1155/2004/765753

Nakashima, K., Zhou, X., Kunkel, G., Zhang, Z., Deng, J. M., Behringer, R. R., et al. (2002). The novel zinc finger-containing transcription factor Osterix is required for osteoblast differentiation and bone formation. *Cell* 108, 17–29. doi: 10.1016/S0092-8674(01)00622-5

Ni, G. X., Yao, Z. P., Huang, G. T., Liu, W. G., and Lu, W. W. (2011). The effect of strontium incorporation in hydroxyapatite on osteoblasts *in vitro*. *J. Mater. Sci. Mater. Med.* 22, 961–967. doi: 10.1007/s10856-011-4264-0

Noshi, T., Yoshikawa, T., Dohi, Y., Ikeuchi, M., Horiuchi, K., Ichijima, K., et al. (2001). Recombinant human bone morphogenetic protein-2 potentiates the *in vivo* osteogenic ability of marrow/hydroxyapatite composites. *Artif. Organs.* 25, 201–8. doi: 10.1046/j.1525-1594.2001.025003201.x

Pan, H. B., Li, Z. Y., Lam, W. M., Wong, J. C., Darvell, B. W., Luk, K. D. K., et al. (2009). Solubility of strontium-substituted apatite by solid titration. *Acta Biomater.* 5, 1678–1685. doi: 10.1016/j.actbio.2008.11.032

Pharmacia and Upjohn Company and Pfizer (2017). *GelFoam - Absorbable gelatin sponge USP*. New York, NY.

Poon, B., Kha, T., Tran, S., and Dass, C. R. (2016). Bone morphogenetic protein-2 and bone therapy: Successes and pitfalls. *J. Pharm. Pharmacol.* 68, 139–147. doi: 10.1111/jphp.12506

Provot, S., and Schipani, E. (2005). Molecular mechanisms of endochondral bone development. *Biochem. Biophys. Res. Commun.* 328, 658–65. doi: 10.1016/j.bbrc.2004.11.068

Qiu, K., Zhao, X. J., Wan, C. X., Zhao, C. S., and Chen, Y. W. (2006). Effect of strontium ions on the growth of ROS17/2.8 cells on porous calcium polyphosphate scaffolds. *Biomaterials* 27, 1277–86. doi: 10.1016/j.biomaterials.2005.08.006

Querido, W., Rossi, A. L., and Farina, M. (2016). The effects of strontium on bone mineral: a review on current knowledge and microanalytical approaches. *Micron* 80, 122–34. doi: 10.1016/j.micron.2015.10.006

Rohanizadeh, R., Swain, M. V., and Mason, R. S. (2008). Gelatin sponges (Gelfoam®) as a scaffold for osteoblasts. *J. Mater. Sci. Mater. Med.* 19, 1173–1182. doi: 10.1007/s10856-007-3154-y

Rohnke, M., Kokesch-Himmelreich, J., Wenisch, S., Schumacher, M., Gelinsky, M., Bernhardt, A., et al. (2016). Strontium substitution in apatitic CaP cements effectively attenuates osteoclastic resorption but does not inhibit osteoclastogenesis. *Acta Biomater.* 37, 184–194. doi: 10.1016/j.actbio.2016.04.016

Rutkovskiy, A., Stensløkken, K.-O., and Vaage, I. J. (2016). Osteoblast differentiation at a glance. *Med. Sci. Monit. Basic Res.* 22, 95–106. doi: 10.12659/MSMBR.901142

Sage, J., Leblanc-Noblesse, E., Nizard, C., Sasaki, T., Schnebert, S., Perrier, E., et al. (2012). Cleavage of nidogen-1 by cathepsin S impairs its binding to basement membrane partners. *PLoS ONE* 7:e43494. doi: 10.1371/journal.pone.0043494

Saidak, Z., and Marie, P. J. (2012). Strontium signaling: molecular mechanisms and therapeutic implications in osteoporosis. *Pharmacol. Ther.* 136, 216–226. doi: 10.1016/j.pharmthera.2012.07.009

Santo, V. E., Gomes, M. E., Mano, J. F., and Reis, R. L. (2013). Controlled release strategies for bone, cartilage, and osteochondral engineering-part i: recapitulation of native tissue healing and variables for the design of delivery systems. *Tissue Eng. B Rev.* 19, 308–326. doi: 10.1089/ten.teb.2012.0138

Scheinpflug, J., Pfeiffenberger, M., Damerau, A., Schwarz, F., Textor, M., Lang, A., et al. (2018). Journey into bone models: a review. *Genes (Basel).* 9:247. doi: 10.3390/genes9050247

Schmittgen, T. D., and Livak, K. J. (2008). Analyzing real-time PCR data by the comparative C(T) method. *Nat. Protoc.* 3, 1101–8. doi: 10.1038/nprot.2008.73

Schumacher, M., Lode, A., Helth, A., and Gelinsky, M. (2013). A novel strontium(II)-modified calcium phosphate bone cement stimulates human-bone-marrow-derived mesenchymal stem cell proliferation and osteogenic differentiation *in vitro*. *Acta Biomater.* 9, 9547–9557. doi: 10.1016/j.actbio.2013.07.027

Sozen, T., Ozisik, L., and Calik Basaran, N. (2017). An overview and management of osteoporosis. *Eur. J. Rheumatol.* 4, 46–56. doi: 10.5152/eurjrheum.2016.048

Takaoka, S., Yamaguchi, T., Yano, S., Yamauchi, M., and Sugimoto, T. (2010). The calcium-sensing receptor (CaR) is involved in strontium ranelate-induced osteoblast differentiation and mineralization. *Horm. Metab. Res.* 42, 627–631. doi: 10.1055/s-0030-1255091

Tian, A., Zhai, J., Peng, Y., Zhang, L., Teng, M., Liao, J., et al. (2014). Osteoblast response to titanium surfaces coated with strontium ranelate-loaded chitosan film. *Int. J. Oral Maxillofac. Implants* 29, 1446–1453. doi: 10.11607/jomi.3806

Urist, M. R., Huo, Y. K., Brownell, A. G., Hohl, W. M., Buyske, J., Lietze, A., et al. (1984). Purification of bovine bone morphogenetic protein by hydroxyapatite chromatography. *Proc. Natl. Acad. Sci. U. S. A.* 81, 371–375. doi: 10.1073/pnas.81.2.371

Visai, L., Cristofaro, F., Corsetto, P. A., Campi, G., Cedola, A., Pascucci, B., et al. (2017). Heterogeneous and self-organizing mineralization of bone matrix promoted by hydroxyapatite nanoparticles. *Nanoscale* 9, 17274–17283. doi: 10.1039/C7NR05013E

Wang, E. A., Rosen, V., D'Alessandro, J. S., Bauduy, M., Cordes, P., Harada, T., et al. (1990). Recombinant human bone morphogenetic protein induces bone formation. *Proc. Natl. Acad. Sci. U. S. A.* 87, 2220–2224. doi: 10.1073/pnas.87.6.2220

Wei, P., Jing, W., Yuan, Z., Huang, Y., Guan, B., Zhang, W., et al. (2019). Vancomycin- and strontium-loaded microspheres with multifunctional activities against bacteria, in angiogenesis, and in osteogenesis for enhancing infected bone regeneration. *ACS Appl. Mater. Interfaces* 11, 30596–30609. doi: 10.1021/acsami.9b10219

Woo, E. J. (2013). Adverse events after recombinant human BMP2 in nonspinal orthopaedic procedures general. *Clin. Orthop. Relat. Res.* 471, 1707–1711. doi: 10.1007/s11999-012-2684-x

Wrobel, E., Leszczynska, J., and Brzoska, E. (2016). The characteristics of human bone-derived cells (HBDCS) during osteogenesis *in vitro*. *Cell. Mol. Biol. Lett.* 21, 1–15. doi: 10.1186/s11658-016-0027-8

Yang, F., Yang, D., Tu, J., Zheng, Q., Cai, L., and Wang, L. (2011). Strontium enhances osteogenic differentiation of mesenchymal stem cells and *in vivo* bone formation by activating Wnt/catenin signaling. *Stem Cell.* 29, 981–991. doi: 10.1002/stem.646